# Spherical Harmonic Representation of Energetic Neutral Atom Flux Components Observed by IBEX

P. Swaczyna[1,*], M. A. Dayeh[2,3], E. J. Zirnstein[1]

[1]Department of Astrophysical Sciences, Princeton University, Princeton, NJ 08544, USA
[2]Southwest Research Institute, San Antonio, TX 78238, USA
[3]Department of Physics and Astronomy, University of Texas at San Antonio, San Antonio, TX 78249, USA

**Abstract**

The Interstellar Boundary Explorer (IBEX) images the heliosphere by observing energetic neutral atoms (ENAs). The IBEX-Hi instrument onboard IBEX provides full-sky maps of ENA fluxes produced in the heliosphere and very local interstellar medium (VLISM) through charge exchange of suprathermal ions with interstellar neutral atoms. The first IBEX-Hi results showed that in addition to the anticipated globally distributed flux (GDF), a narrow and bright emission from a circular region in the sky, dubbed the IBEX ribbon, is visible in all energy steps. While the GDF is mainly produced in the inner heliosheath, ample evidence indicates that the ribbon forms outside the heliopause in the regions where the interstellar magnetic field is perpendicular to the lines of sight. The IBEX maps produced by the mission team distribute the observations into 6°×6° rectangle pixels in ecliptic coordinates. The overlap of the GDF and ribbon components complicates qualitative analyses of each source. Here, we find the spherical harmonic representation of the IBEX maps, separating the GDF and ribbon components. This representation describes the ENA flux components in the sky without relying on any pixelization scheme. Using this separation, we discuss the temporal evolution of each component over the solar cycle. We find that the GDF is characterized by larger spatial scale structures than the ribbon. However, we identify two isolated, small-scale signals in the GDF region that require further study.

## 1. Introduction

Imaging the heliosphere through observations of energetic neutral atom (ENA) emission is an important tool for understanding the global structure of the interaction of the solar wind with the very local interstellar medium (VLISM) and its evolution over solar cycles (Gruntman 1997). The Interstellar Boundary Explorer (IBEX) is the first mission to focus on observations of heliospheric ENAs (McComas et al. 2009a). The spacecraft includes two ENA imagers: IBEX-Hi covering energies from 0.5 to 6 keV (Funsten et al. 2009) and IBEX-Lo observing neutral atoms from 10 eV to 2 keV (Fuselier et al. 2009b). Most ENA observations used to analyze the global heliosphere are collected with IBEX-Hi. The first full-sky maps of ENA flux showed a narrow circular emission of ENAs called the IBEX ribbon, clearly visible in addition to the globally distributed flux (GDF) in all IBEX-Hi energy steps (Fuselier et al. 2009a; McComas et al. 2009b, 2012, 2014a, 2017, 2020; Schwadron et al. 2009).

Prior to IBEX, the understanding was that heliospheric ENAs are primarily created from suprathermal ions in the inner heliosheath. Therefore, modeled ENA flux maps showed broad structures connected to the latitudinal configuration of the solar wind and heliospheric nose-tail asymmetry (Heerikhuisen et al. 2008, 2009). The ribbon, discovered as a narrow structure stretching over most of the sky, is not clearly organized by the interstellar flow direction or heliographic latitude. The first ribbon analyses revealed, however, that it is observed in directions where the lines of sight are perpendicular to the interstellar magnetic field (Schwadron et al. 2009). Soon after, more than a dozen hypotheses were formulated to explain the IBEX ribbon, placing the source region of the ribbon ENAs from the termination shock to a distant boundary layer in the interstellar medium (see review by McComas et al. 2014b). Detailed analyses of the IBEX ribbon indicate that it is likely created in the secondary ENA mechanism (Chalov et al. 2010; Heerikhuisen et al. 2010; Schwadron & McComas 2013; Zirnstein et al. 2013). In this mechanism, ions in the solar wind are neutralized and expand as neutral solar wind to distances beyond the heliopause, where two subsequent charge exchanges produce secondary ENAs. Soma secondary ENAs travel back to the proximity of the Sun and are visible as the IBEX ribbon.

[*] Corresponding author (swaczyna@princeton.edu)

In each energy step, the ribbon follows a circle in the sky with a radius of ~75° centered at a point close to the B-V plane defined by the directions of the interstellar flow and magnetic field (Funsten et al. 2013). The centers of the ribbon in subsequent ESA steps show a systematic progression, mainly in the direction perpendicular to the B-V plane. This progression is related to the structure of the solar wind (Swaczyna et al. 2016b) and is present throughout the solar cycle (Dayeh et al. 2019). The geometry of the ribbon is also critical to determine the interstellar magnetic field in the VLISM (Zirnstein et al. 2016) and to understand the details of the secondary ENA mechanism (Zirnstein et al. 2015, 2019a, 2019b, 2021a).

Analyses of the IBEX ribbon are difficult because of the overlapping GDF and ribbon components. Several methodologies have been developed to perform a global separation of these two signals assuming either a functional form of the ribbon profile (Funsten et al. 2013, 2015; Dayeh et al. 2019, 2023a; Swaczyna et al. 2016a; Reisenfeld et al. 2021) or by interpolating the GDF inside the ribbon region (Schwadron et al. 2011, 2014, 2018). Beesley et al. (2023) developed statistical methods for separating these sources applied to higher-resolution maps generated by Osthus et al. (2022). Furthermore, Swaczyna et al. (2022a, hereafter Paper I) proposed an approach in which spherical harmonics are used as a basis to represent the GDF, enabling estimation of the ribbon flux as the difference between the IBEX observed maps and the GDF reproduced from the spherical harmonic decomposition. In this paper, we further develop this methodology by extending the spherical harmonic representation to both components.

## 2. Methodology

The spherical harmonic decomposition of ENA flux maps presented in this paper bases on the methodology presented in Paper I. However, we implement several changes that we discuss in this section. First, we do not calculate the spherical harmonic coefficients but find them by minimizing a least-squares term (Section 2.1). Furthermore, we introduce a regularization term to eliminate overfitting in underconstrained regions (Section 2.2). Our analysis includes higher degree spherical harmonics to properly reconstruct narrow structures observed, especially within the IBEX ribbon region (Section 2.3). We also modify the ribbon mask definition used to estimate the region with the ribbon component (Section 2.4). Finally, in Section 2.5, we discuss the uncertainty analysis of the separated components.

### *2.1 Least-squares Minimization*

Standard maps provided in the IBEX data releases are represented by ENA flux values in 6°×6° pixels defined on a regular grid in the ecliptic coordinates. This pixelization scheme allows for an approximately uniform distribution of exposure times because data collected by IBEX for a spin axis pointing exactly in the ecliptic plane are distributed into vertical strips at longitudes ±90° from the spin axis pointing longitude.

The regular repointing provides full sky maps on a half-year cadence but the observations are often separated into maps constructed from one year of observations in the ram and anti-ram hemispheres. We focus on yearly ram maps in this study, but the procedure can be applied to other types of IBEX maps. Map files are organized so that the first row represents the pixels near the south ecliptic pole (centered at latitude $\beta = -87°$) for increasing longitudes $\lambda = 3°, 9°, \dots, 357°$. The subsequent rows provide fluxes in pixels at higher latitudes. However, for the purpose of the study, we organize each IBEX map into a single vector in which the data from all rows are joined together. Therefore, for each map, we construct a vector of observed ENA flux values: $\boldsymbol{j} = \{j_k\}_{k=1,\dots,N_{\text{pix}}}$ in $N_{\text{pix}} = 1800$ pixels and a diagonal covariance matrix $\mathbf{V} = \text{diag}\left(\{\sigma_k^2\}_{k=1,\dots,N_{\text{pix}}}\right)$ with the flux variances at the matrix diagonal. The data gaps can be included with any value in the vector $\boldsymbol{j}$. However, the inverted covariance matrix $\mathbf{V}^{-1}$ must have zeros at the corresponding diagonal positions.

As in Paper I, we use the real representation of spherical harmonics $Y_{\ell m}(\theta, \phi)$, where $\ell$ and $m$ are the degree and order of the spherical harmonic, while $\theta$ and $\phi$ are the azimuth and polar angle. We transform the ecliptic latitudes to their co-latitudes, which are polar angles. There are $2\ell + 1$ orders of spherical harmonics $m = -\ell, -\ell + 1, \dots, \ell$ for each degree $\ell$. Therefore, we have $(\ell_{\max} + 1)^2$ spherical harmonics for all degrees $\ell \leq \ell_{\max}$. In this study, we organize them into a vector ordered by degree and order. Let $y_{k,\ell m}$ denotes the average of the spherical harmonic $Y_{\ell m}(\theta, \phi)$ over the solid angle corresponding to pixel $k$ (see Swaczyna et al. 2022a for details). Next, we define a vector $\boldsymbol{y}_k = \{y_{k,\ell m}\}_{(\ell m):\ell \leq \ell_{\max}}$ of these average values for each pixel $k$ and all spherical harmonics with $\ell \leq \ell_{\max}$. Finally, these vectors for all pixels can be organized into rows of a $N_{\text{pix}} \times (\ell_{\max} + 1)^2$ matrix $\mathbf{Y} = \{\boldsymbol{y}_k\}_{k=1,\dots,N_{\text{pix}}}$.

Here, we want to find a vector of spherical harmonic coefficients $\boldsymbol{c} = \{c_{\ell m}\}_{(\ell m):\ell \leq \ell_{\max}}$ that minimizes the following least-square sum:

$$\chi^2_{\text{LS}}(\{c_{\ell m}\}_{(\ell m):\ell \leq \ell_{\max}}) = \sum_{k=1}^{N_{\text{pix}}} \frac{\left(\sum_{(\ell m)} y_{k,\ell m} c_{\ell m} - j_k\right)^2}{\sigma_k^2}. \quad (1)$$

This minimization allows us to find the coefficients that minimize the residual fluxes weighted by their uncertainties. Note that this approach differs from the one introduced in Paper I because we explicitly include uncertainties in this procedure. Using the terminology presented above, we can rewrite Equation (1) in a condensed form:

$$\chi^2_{\text{LS}}(\boldsymbol{c}) = (\mathbf{Y}\boldsymbol{c} - \boldsymbol{j})^{\text{T}} \mathbf{V}^{-1} (\mathbf{Y}\boldsymbol{c} - \boldsymbol{j}). \quad (2)$$

Note that the data gaps for which the inverted covariance is zero are effectively excluded from this sum. The minimization of the above expression represents the best-fit spherical harmonic representation of the ENA map represented in vector $\boldsymbol{j}$. However, high-degree spherical harmonics, used in our study, can sometimes represent features smaller than the data gaps, which this expression would not constrain. Therefore, to minimize the impact of these gaps, we use a regularization term, as described in the next section.

*2.2 Tikhonov Regularization and L-curve*

The spherical harmonic decomposition of a single IBEX map is described by $(\ell_{\max}+1)^2$ coefficients. The minimization of the term given in Equation (1) is clearly ill-posed (underconstrained) if $\ell_{\max} > \sqrt{N_{\text{pix}}} - 1 \approx 41.4$. However, because the IBEX maps frequently include gaps, we need to introduce a regularization term in the minimization even though we are not using as many degrees of spherical harmonics (see Section 2.3). Moreover, the ribbon masking procedure discussed further in the paper also requires regularization.

In our analysis, we implement Tikhonov regularization (see, e.g., Tikhonov et al. 1995; Calvetti et al. 2000) in the following form:

$$\chi^2_{\text{reg}} = \boldsymbol{c}^{\text{T}}\mathbf{R}\boldsymbol{c}, \tag{3}$$

where $\mathbf{R}$ is a regularization matrix. The selection of the regularization matrix depends on the regularization goal. In our case, we want to minimize gradients, i.e., we assume the flux does not significantly change over the gaps. Metzler & Pail (2005) showed that to minimize the absolute gradients of the spherical harmonic decomposition integrated over the sphere, one needs to implement the following diagonal regularization matrix:

$$\mathbf{R} = \left\{\ell_i(\ell_i+1)\delta_{\ell_i,\ell_j}\delta_{m_i,m_j}\right\}_{i,j}, \tag{4}$$

where $i$ and $j$ enumerate the spherical harmonics up to the degree of $\ell_{\max}$, and $\delta_{i,j}$ is the Kronecker delta. The regularization term aims to penalize high-degree spherical harmonics, which describe smaller-scale structures for which the spatial scale is larger.

The minimization of two terms given in Equations (2) and (3) cannot be performed independently but needs to be implemented as a single minimization requirement. Therefore, the joint minimization is typically considered by a sum of these two terms with an unknown regularization parameter $\alpha$:

$$\chi^2 = \chi^2_{\text{LS}} + \alpha\chi^2_{\text{reg}} = (\mathbf{Y}\boldsymbol{c} - \boldsymbol{j})^{\text{T}}\mathbf{V}^{-1}(\mathbf{Y}\boldsymbol{c} - \boldsymbol{j}) + \alpha\boldsymbol{c}^{\text{T}}\mathbf{R}\boldsymbol{c} \tag{5}$$

Equation (5) minimization can be performed analytically because our model is linear. The coefficients $\hat{\boldsymbol{c}}(\alpha)$ that minimize this equation are (e.g., Metzler & Pail 2005):

$$\hat{\boldsymbol{c}}(\alpha) = \left(\mathbf{Y}^{\text{T}}\mathbf{V}^{-1}\mathbf{Y} + \alpha\mathbf{R}\right)^{-1}\mathbf{Y}^{\text{T}}\mathbf{V}^{-1}\boldsymbol{j}. \tag{6}$$

The optimal regularization parameter $\alpha$ can be obtained from analysis of the trajectory of the logarithms of the minimization terms $(\log\chi^2_{\text{LS}}, \log\chi^2_{\text{reg}})$ obtained from the minimization of Equation (5) for various values of the regularization parameters, known as the L-curve (e.g., Hansen & O'Leary 1993; Calvetti et al. 2000). Figure 1 presents this trajectory for the ram-only 2016 map for energy step 1.7 keV. This map includes significant data gaps; thus, it is a good example to show the role of regularization. The L-curve name comes from its shape, as the trajectory includes a corner where the change of the logarithms is the smallest relative to each other. This point is quantified based on the point of the highest curvature. Near this point, the changes of the optimal $\chi^2_{\text{LS}}$ and $\chi^2_{\text{reg}}$ are the smallest with respect to each other. In other words, in other parts of the curve, a slight change in one of these results in a significant change in the other. The bottom panels of Figure 1 present the maps reconstructed from the spherical harmonic representation for $\ell_{\max} = 22$ for three regularization parameter values. If the regularization parameter is lower than the optimal, the map reconstructed from the spherical harmonic reconstruction show structure inside the data gap region (e.g., a bright spot near the center of the map in the bottom left panel of Figure 1). If the regularization parameter is larger than optimal, the reconstructed map is smoothed and does not preserve the small-scale features of the original map.

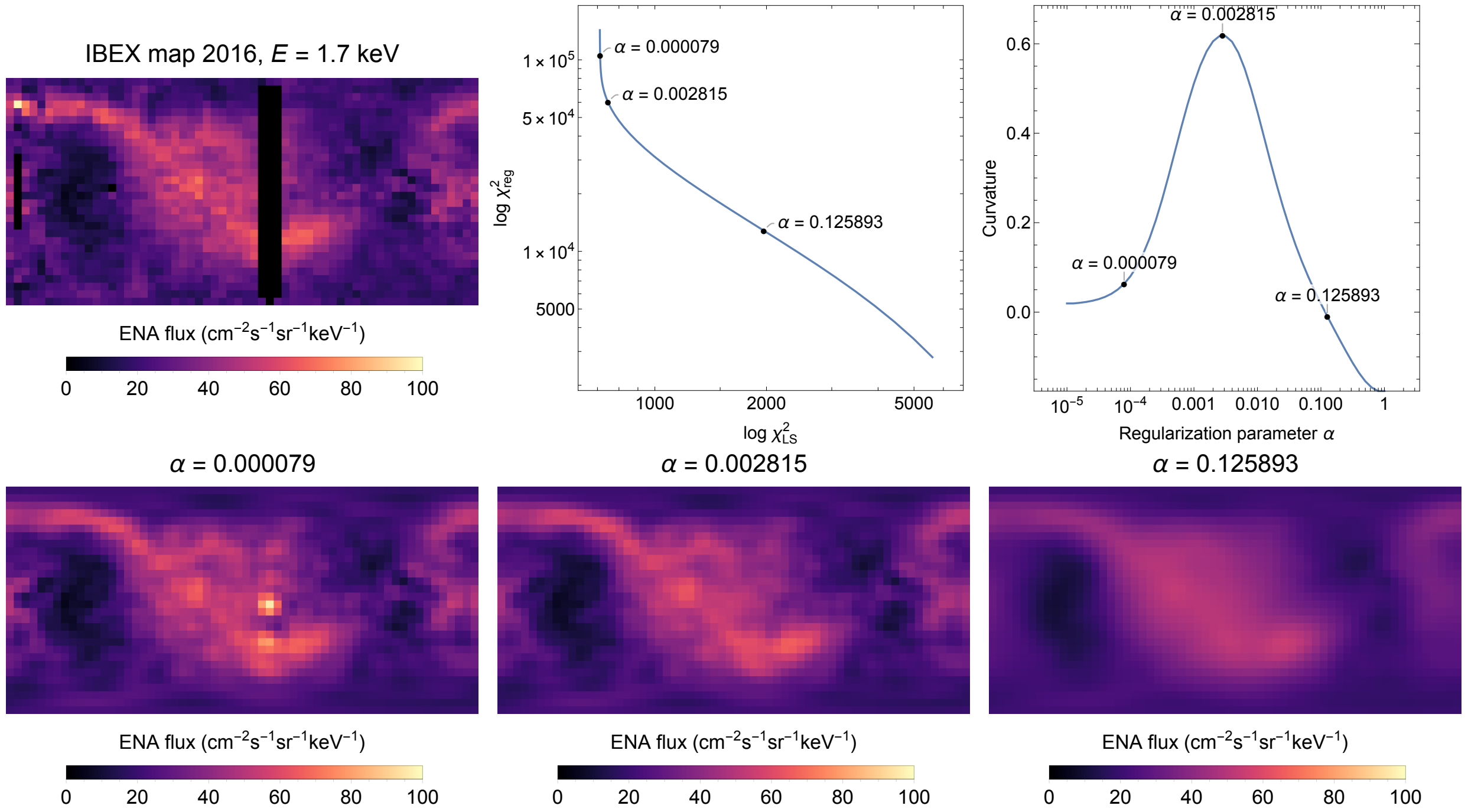

**Figure 1.** Role of the regularization for spherical harmonic decomposition shown for the IBEX map at energy step 1.7 keV observed in 2016, which includes a significant gap in the data (top left panel). The top middle and right panels show the L-curve and the curvature as a function of the regularization parameter, respectively. The three bottom panels show the reconstructed maps for three values of the regularization parameter: $\alpha = 0.000079, 0.002815$, and $0.125893$. The middle panel shows the result for optimal regularization. The maps shown here and in other figures in this paper use the equirectangular projection in ecliptic coordinates where the top and bottom edges correspond to the north and south ecliptic poles, respectively. The left and right edges correspond to ecliptic longitude of 72°.

### *2.3 Maximum Degree of Spherical Harmonics $\ell_{max}$*

Most of the globally distributed flux can be reconstructed from spherical harmonics up to the degree of 3 (see Paper I). However, here we provide an alternative representation of all features, including the IBEX ribbon and other small-scale structures. Therefore, we significantly increase the maximum degree $\ell_{\max}$. In this section, we justify that $\ell_{\max} = 22$ allows for sufficient representation of the IBEX data presented in the IBEX maps.

Our first argument utilizes decomposition to a much higher degree of $\ell_{\max} = 30$ for the time-combined maps in IBEX Data Release #16[†], covering data collected between 2009 and 2019. The time-combined maps have the lowest uncertainties. Therefore, we use them to estimate the required degree that allows for representation of all small-scale structures. We perform the fitting described in the sections above to find the coefficients $c_{\ell m}$ and calculate the related uncertainties $\delta c_{\ell m}$ (see Section 2.5). For each degree of spherical harmonics, we calculate the mean statistical significance over all orders of spherical harmonics:

$$\sigma_\ell = \frac{1}{2\ell+1} \sum_{m=-\ell}^{\ell} \left| \frac{c_{\ell m}}{\delta c_{\ell m}} \right|, \tag{7}$$

which are shown for all energy steps in the top panel of Figure 2. This figure shows that the highest-degree statistically significant coefficients ($\sigma_\ell > 1$) depend on the energy step and range from $\ell = 17$ (for the energy step 0.7 keV) to $\ell = 22$ (for 1.7 keV). The contribution from spherical harmonics for degrees above this limit on average over the entire sky is not statistically significant.

[†] https://ibex.princeton.edu/DataRelease16

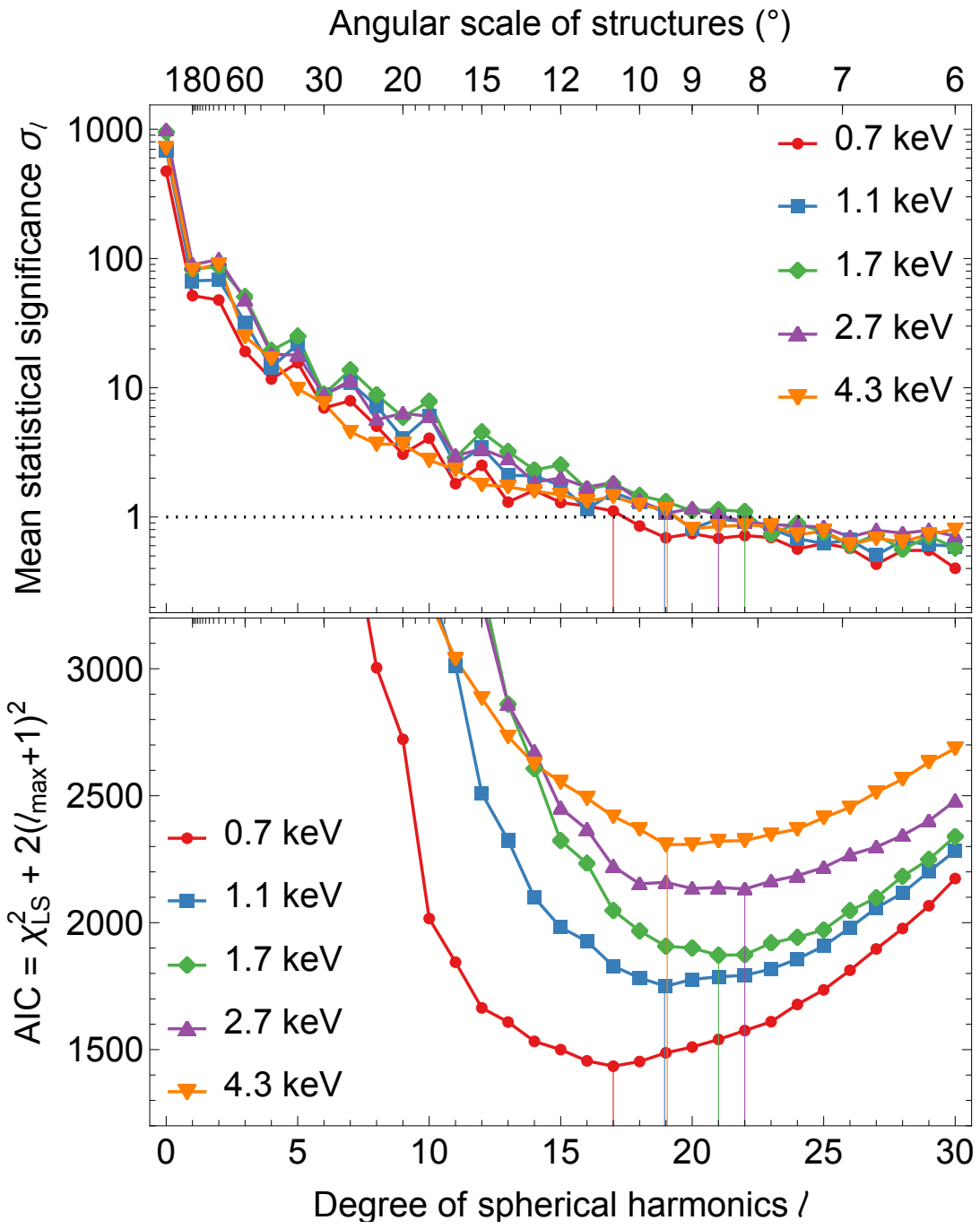

**Figure 2.** Optimization of the maximum degree of spherical harmonics. The top panel shows the mean statistical significance of spherical harmonic contributions for all orders as a function of the degree. The bottom panel shows the AIC used to select among models with various numbers of parameters. The optimal maximum degree ranges from 17 to 22. The top scale shows the angular size of structures that can be reconstructed.

Additionally, we consider the Akaike Information Criterion (AIC, Akaike 1974), calculated as

$$\mathrm{AIC} = \chi^2_{\mathrm{LS,min}} + 2(\ell_{\mathrm{max}} + 1)^2, \tag{8}$$

which adds a penalty term proportional to the number of the model parameters. While the minimum $\chi^2_{\mathrm{LS,min}}$ always decreases as the highest degree $\ell_{\mathrm{max}}$ increases because additional parameters allow for a better representation of the original map, the penalty term allows us to find the degree that minimizes the AIC. The bottom panel of Figure 2 presents the AIC value as a function of the maximum degree. The minimum AIC is obtained for $\ell_{\mathrm{max}}$ from 17 (0.7 keV) to 22 (2.7 keV).

Generally, spherical harmonics up to some degree $\ell$ can only represent structures with characteristic sizes of the order of $180°/\ell$. Therefore, the IBEX-Hi ~6°×6° field-of-view limits the smallest structures that can be observed to about $\ell \approx 30$. However, the pixelization scheme decreases the effective resolution. Because the angular pixel size changes from the ecliptic plane to the poles, the effective resolution differs, but one can make an approximation that this leads to an effective resolution of $\sim((6°)^2 + (6°)^2)^{0.5} \approx 8.5°$, which corresponds to $\ell \approx 21$.

Figure 3 compares the reconstructed full maps for $E = 1.7$ keV and $\ell_{\max} = 14$, 22, and 30, together with the normalized residuals calculated as differences between the original and reconstructed maps divided by the uncertainty of the original fluxes. The residuals for $\ell_{\max} = 14$ show systematic patterns that appear to be in the regions of the most substantial gradients in the IBEX ribbon. Therefore, we need more spherical harmonics to describe the ribbon. On the other hand, the result for $\ell_{\max} = 30$ shows a significant reduction in the relative residuals. In general, the variation is much smaller than expected from the uncertainties, which means that a significant amount of statistical variation from the original map is reflected in the spherical harmonic decomposition. Moreover, the magnitude of the relative residuals increases toward the poles because angular distances between pixels are smaller, and thus the spherical harmonics cannot reconstruct this variation. This effect is, to some extent, visible for $\ell_{\max} = 22$. Nevertheless, we decide to use $\ell_{\max} = 22$ for our study. While some small-scale structures may be a result of statistical fluctuations, it is important that we can represent the ribbon structure fully. This decomposition may be generally used for filtration of the IBEX map, but it is not our goal in this study.

### *2.4 Ribbon Mask and Spherical Harmonic Decomposition of the GDF*

The main characteristic that led to the discovery of the IBEX ribbon was that it is limited to a ring-shaped region of the sky (Funsten et al. 2013; Dayeh et al. 2019). While some ENA emission originating in the outer heliosheath may form broader structures (Zirnstein et al. 2019a), we cannot separate them from the GDF as they have similar geometric structures. Therefore, these broad structures are not considered part of the ribbon. In other words, our working definition of the ribbon focuses on the geometric property of the ribbon, i.e., that it is a narrow circular structure seen in the IBEX maps, rather than on the source region of the ENAs.

We need to find a set of pixels in IBEX maps that include the ribbon component to mask them out for the derivation of the GDF. In Paper I, an iterative procedure was employed to select pixels where the magnitude of the residual flux calculated between the original IBEX map and the GDF is statistically high. The pixels with high sum values were classified into the ribbon mask. However, this criterion tends to provide a ribbon mask which is narrower where the ribbon is weaker compared to the underlying GDF because they result in lower statistical significance. However, lower ribbon fluxes do not mean that the ribbon is narrower. This effect can be noted in Figure 1 of Paper I in pixels near the nose and ecliptic plane, especially for $\ell_{max} \geq 3$. As we use higher-degree spherical harmonics, we cannot follow the same procedure here. Therefore, we introduce a new method to find the ribbon based on the observation that it is limited to a ring-shaped region. Moreover, we define a separate mask for each energy step because the ribbon position depends strongly on energy but only slightly changes over time (Funsten et al. 2013; Swaczyna et al. 2016b; Dayeh et al. 2019; Zirnstein et al. 2023). Therefore, we use one mask for all years. For the derivation of the mask, we use the combined data from 2009 to 2011. We do not use later observations because the ribbon fluxes drop significantly, reducing the contrast of the ribbon to the GDF emissions, especially in the highest energy step.

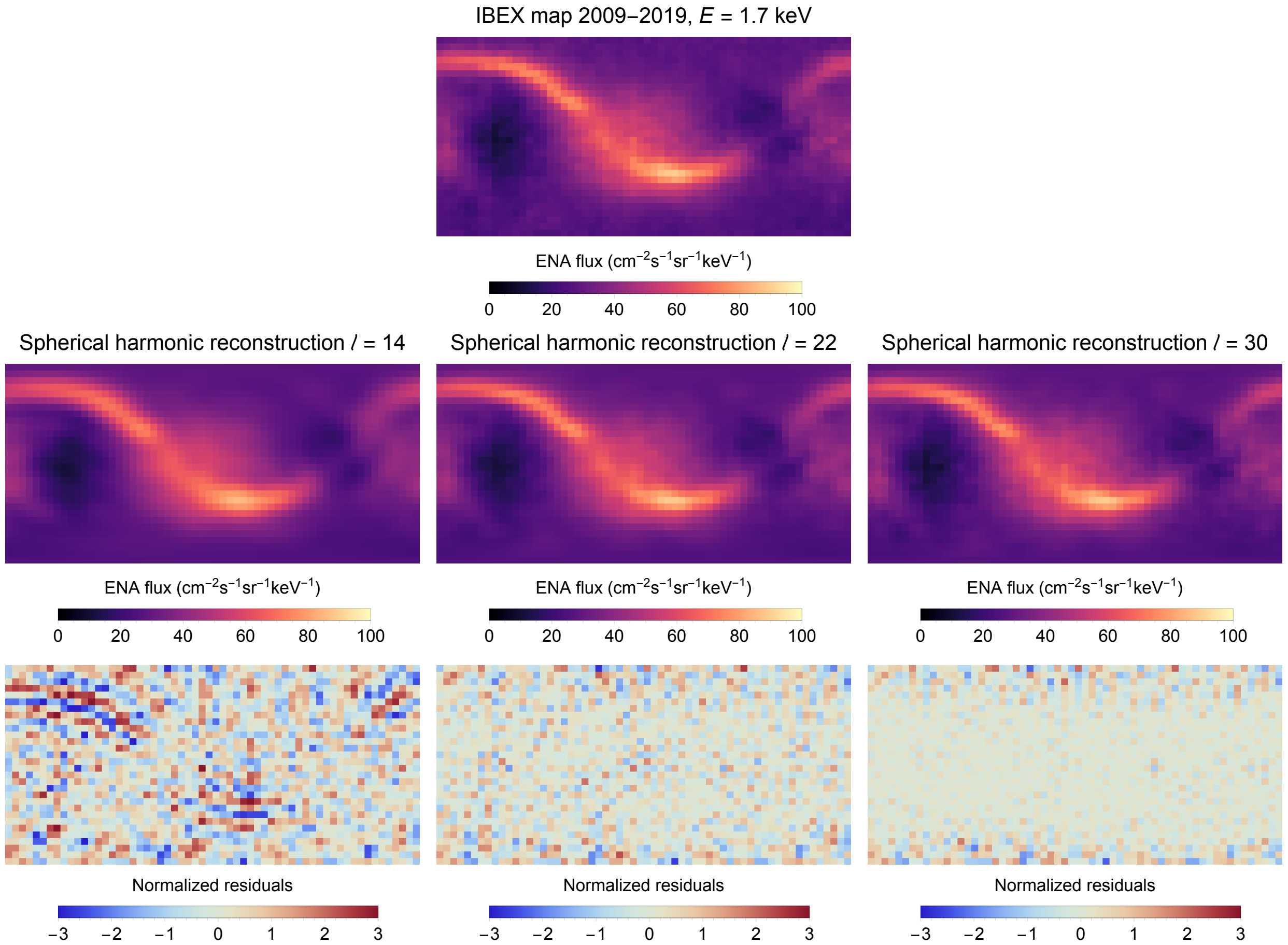

**Figure 3.** Spherical harmonic reconstruction of the time-combined IBEX map at energy step 1.7 keV. The top panel shows the original map, and the middle panels show the reconstruction for the maximum degree of spherical harmonics $\ell_{\max} = 14$, 22, and 30 (left to right). The bottom panels show the residual signal normalized by the original map uncertainties.

The ribbon masks considered for this study are defined by center $(\lambda, \beta)$, radius $r$, and width $w$. The ribbon mask is generally defined as pixels for which the angular distance between the pixels' centers and the mask's center is within the range $[r - w/2, r + w/2]$. From the perspective of the GDF, the mask should be as narrow as possible to reconstruct as many small-scale structures of the GDF as possible. On the other hand, it needs to be wide enough to encompass the ribbon region. Spherical harmonics up to some degree $\ell$ describe structures that have sizes not smaller than $180°/\ell$. We decide that we want to be able to reconstruct the GDF structures up to the degree of $\ell = 4$, and thus we select the width of $w = 180°/4 = 45°$ for our study. While Paper I showed that spherical harmonics could capture most structures up to $\ell = 3$, we want to describe smaller-scale structures in this study, and the power spectrum suggests that the spherical harmonics for $\ell = 4$ contribute significantly to the GDF in some energy steps (see Section 4.1). Nevertheless, we will verify our results for the choice of the full width corresponding to $\ell = 3$ and 5 ($w = 60°$ and 36°) to illustrate the role of this parameter. While the selected mask sizes correspond to low-degree spherical harmonics, we estimate the decomposition of the GDF with the same $\ell_{\max} = 22$ as used in the decomposition of the combined flux in Section 2.3. This procedure ensures that the variations outside the ribbon are equally well captured for the combined and GDF-only decompositions.

The ribbon center and radius are refined in an iterative procedure. As the starting parameters, we use the time- and energy-averaged ribbon center and radius obtained by Dayeh et al. (2019). Based on these, we define the pixels inside the ribbon mask. These pixels are excluded from the derivation of the spherical harmonic decomposition of the GDF. Later, the analysis is performed identically to the one presented in Sections 2.1 and 2.2. This removal is achieved by replacing the diagonal elements corresponding to the mask's pixels in the diagonal matrix $\mathbf{V}^{-1}$ with zeros. After the separation, a new ribbon center and radius are calculated using method presented in Appendix A. This method uses the spherical harmonic representation of the ribbon calculated as the difference between the spherical harmonic coefficients obtained for the original map and the map with the ribbon masked. We find the ribbon center position for which the ribbon flux as a function of the distance from this center forms the narrowest structure when averaged over the sky. The radius of the next mask is adopted from the mean distance of the ribbon flux from this center. Finally, we repeat the calculation of the spherical harmonic representation using the new mask.

The iteration is performed until the last mask is identical to one of the masks obtained earlier in the iteration. In some cases, the iteration converges on a single mask, i.e., a mask leads to the ribbon center and radius that defines the same mask. However, the iteration may result in a cycle of ribbon centers. Within each cycle, the ribbon centers and radii are close to each other, much closer than the uncertainty of these parameters. However, these small changes may change at most 4% of pixels included in the mask. Therefore, from the mask in the cycle, we select the one resulting in the narrowest ribbon, as defined in the procedure given in Appendix A.

Figure 4 shows the separation results applied to the ENA flux map at the energy step 1.7 keV for three masks obtained in the above-described process for the assumed width of the ribbon $w = 36°$, 45°, and 60°. The comparison shows that the separation results are similar for these three values, but subtle differences show the width's importance. First, the narrow mask ($w = 36°$) results in a GDF estimate that shows some structures following the edges of the ribbon mask (Figure 4, third column, top panel). This effect is evident in the northern part of the map and results from the fact that some regions with flanks of the ribbon flux remain outside of the mask, and therefore part of the ribbon flux is included in the GDF. On the other hand, the wide mask ($w = 60°$) cannot fully reconstruct the GDF enhancement near the nose, which overlaps partially with the ribbon near the center of the map. This mask extends on the south side beyond the enhancement, so the GDF estimation cannot correctly capture it. The middle mask ($w = 45°$) minimizes these two problems, and we use it further in the analysis.

All GDF maps obtained from the above procedure show smoothing inside the ribbon mask because this part is constrained through the regularization term (Section 2.2), which aims to minimize gradients. Consequently, any analysis of small-scale structures in the GDF should be limited to the regions outside of the mask. The ribbon flux is obtained by subtracting the spherical harmonic decompositions of the total and GDF maps. Consequently, the ribbon estimation may result in negative fluxes, especially outside the ribbon mask. Negative fluxes are not statistically significant, i.e., their absolute values are, on average, smaller than the related uncertainties. Nevertheless, to avoid data selection bias, the pixels with negative values should not be removed from comparisons with models, nor should these values be replaced with zeroes. Both procedures would make the pixels with positive fluxes unbalanced, suggesting small positive background. Nevertheless, comparisons for the ribbon maps can also be limited to the extent of the ribbon mask.

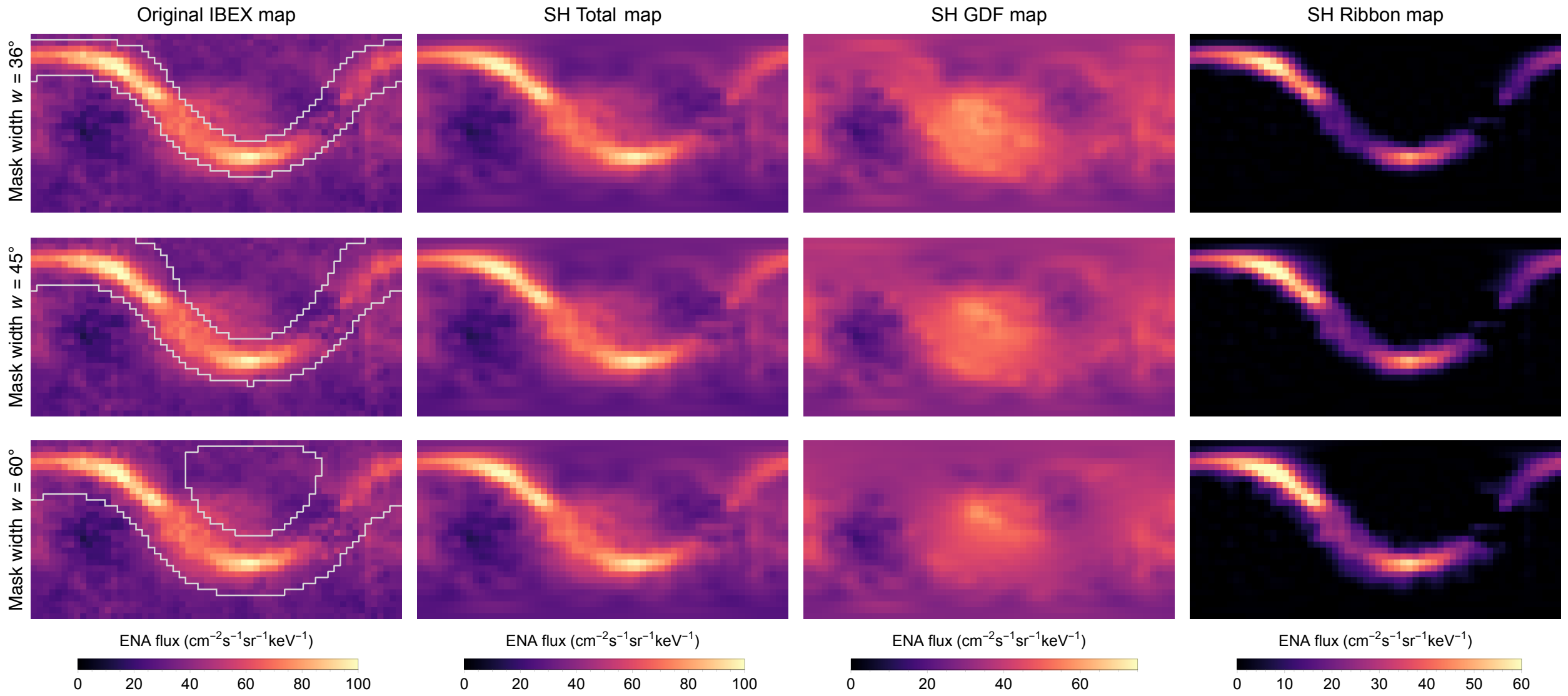

**Figure 4.** Results of the GDF and ribbon separation for combined ENA flux observed in 2009-2011 in energy step 1.7 keV and three possible mask widths of 36°, 45°, and 60° (top to bottom). The maps from left to right show the original IBEX map, with the white outlines showing the extent of the mask, the reconstructed total flux map, the GDF map, and the ribbon map.

### *2.5 Spherical Harmonic Representations and Their Uncertainties*

The spherical harmonic representations of the total and GDF maps are calculated from Equation (6) for the optimal regularization parameter obtained, as discussed in Section 2.2. Writing this equation separately for the total and GDF maps we obtain:

$$\boldsymbol{c}_{\mathrm{T}} = \underbrace{\left(\mathbf{Y}^{\mathrm{T}}\mathbf{V}_{\mathrm{T}}^{-1}\mathbf{Y} + \alpha_{\mathrm{T}}\mathbf{R}\right)^{-1}\mathbf{Y}^{\mathrm{T}}\mathbf{V}_{\mathrm{T}}^{-1}}_{\stackrel{\mathrm{def}}{=}\mathbf{M}_{\mathrm{T}}}\boldsymbol{j}, \tag{10}$$

$$\boldsymbol{c}_{\mathrm{G}} = \underbrace{\left(\mathbf{Y}^{\mathrm{T}}\mathbf{V}_{\mathrm{G}}^{-1}\mathbf{Y} + \alpha_{\mathrm{G}}\mathbf{R}\right)^{-1}\mathbf{Y}^{\mathrm{T}}\mathbf{V}_{\mathrm{G}}^{-1}}_{\stackrel{\mathrm{def}}{=}\mathbf{M}_{\mathrm{G}}}\boldsymbol{j}, \tag{11}$$

where subscripts T and G denote the quantities for the total and GDF maps, respectively. As discussed above, for the total map, the covariance matrix is adopted from the data $\mathbf{V}_{\mathrm{T}} = \mathbf{V}$, which is a diagonal matrix with the flux variance. It is important to note that for the missing data (gaps), the variance is undefined (infinite) even though it is reported in the IBEX data releases as 0. Therefore, the inverted matrix has zeros at the diagonal positions corresponding to the missing data points. Similarly, for the inverted covariance matrix $\mathbf{V}_{\mathrm{G}}$ we replace the values at the positions corresponding to pixels within the ribbon mask with zeros. The regularization parameters $\alpha_{\mathrm{T}}$ and $\alpha_{\mathrm{G}}$ are obtained separately for each map, as discussed in Section 2.2. Finally, the ribbon representation (subscript R) is obtained as the difference of the coefficients:

$$\boldsymbol{c}_{\mathrm{R}} = \boldsymbol{c}_{\mathrm{T}} - \boldsymbol{c}_{\mathrm{G}} = \underbrace{(\mathbf{M}_{\mathrm{T}} - \mathbf{M}_{\mathrm{G}})}_{\stackrel{\mathrm{def}}{=}\mathbf{M}_{\mathrm{R}}}\boldsymbol{j}. \tag{12}$$

Equations (10–12) show that the spherical harmonic representations are a linear combination of the observed ENA fluxes. Therefore, the covariance matrix of these coefficients can be obtained from uncertainty propagation as:

$$\mathbf{W}_a = \mathbf{M}_a \mathbf{V} \mathbf{M}_a^{\mathrm{T}}, \tag{13}$$

where $a$ = T, G, or R.

The spherical harmonic coefficients and their covariance matrix can be transformed into flux values and their covariance matrix on the original IBEX pixelization using simple transformations:

$$\tilde{\boldsymbol{j}}_a = \mathbf{Y} \boldsymbol{c}_a, \tag{14}$$

$$\tilde{\mathbf{V}}_a = \mathbf{Y} \mathbf{W}_a \mathbf{Y}^{\mathrm{T}}. \tag{15}$$

We use tildes to mark that these are reconstructed from the spherical harmonic decomposition. The results of these transformations for the 2016 map and the energy step 1.7 keV are shown in the top and middle rows of Figure 5. This map has a significant data gap near the middle of the map, and thus is a good example to discuss the method's performance. The uncertainty maps show the square root of the diagonal values of the covariance matrix. While for the original map this is complete information, the maps reconstructed from the spherical harmonic representation have spatial correlations that cannot be represented in this figure. We use the same scale for all uncertainties maps to show that the uncertainties are smaller because we effectively use information from neighboring pixels to obtain the reconstructed flux in each of the considered pixels. The methodology uses the information that the structures are described by spherical harmonics up to $\ell_{\max} = 22$. The uncertainty of the ribbon map shows that the uncertainty is the highest inside the ribbon mask and is reduced outside, which is due to the assumption of our analysis that outside of the ribbon, the flux is only from the GDF. However, the GDF map shows that the uncertainty is reduced inside the ribbon mask and is only higher at the edges of the mask. A similar situation is visible in the total map for the data gap regions.

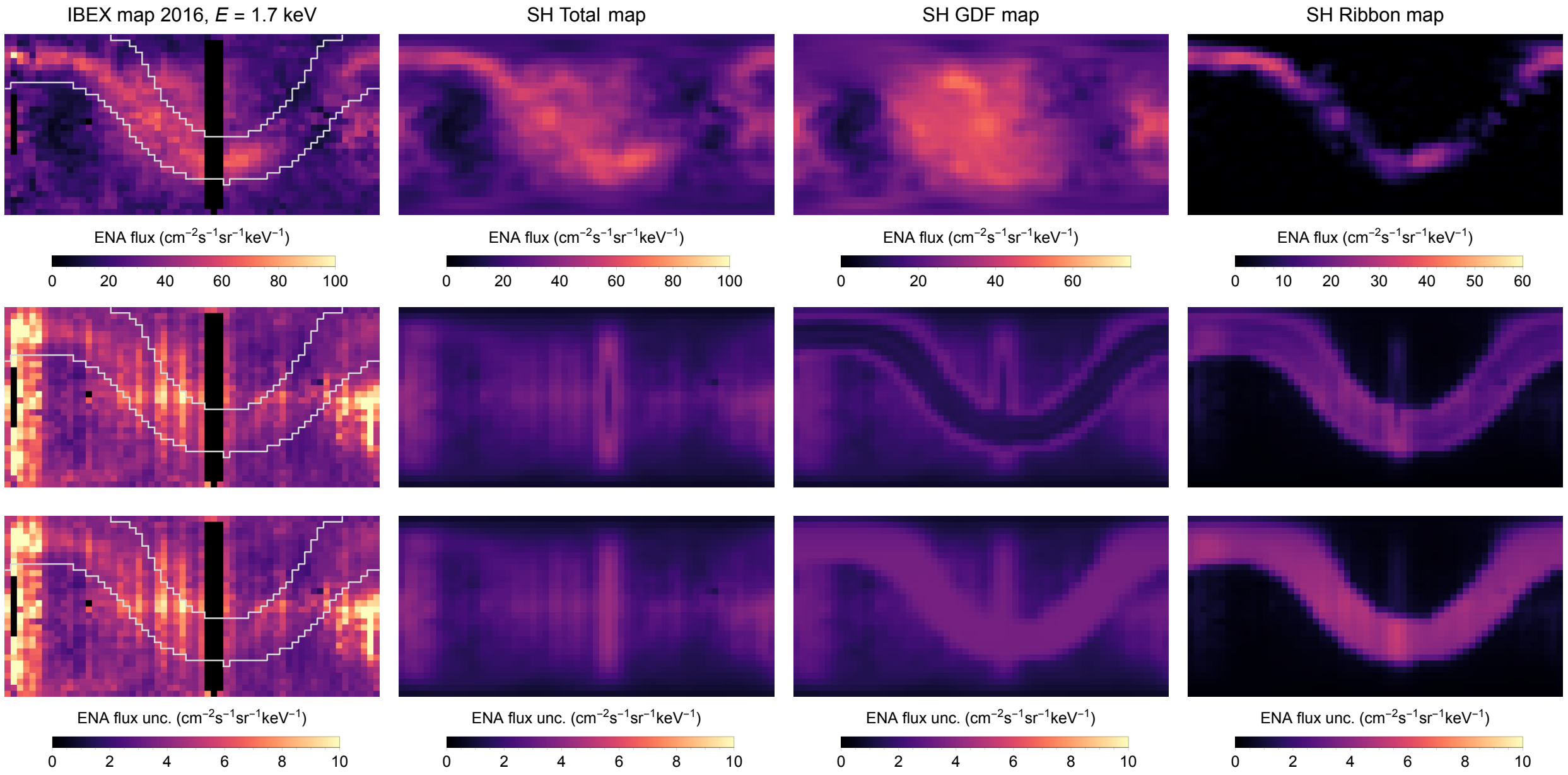

**Figure 5.** Separation for the 2016 map in energy step 1.7 keV. The top panels show the original map and the reconstructed fluxes from the spherical harmonic representations of the total, GDF, and ribbon flux (left to right). The middle and bottom rows show the uncertainties obtained from Equations (13) and (16–17), respectively. The uncertainty map for the original map is the same in both rows.

The underestimation of the uncertainty inside the ribbon mask and data gap regions results from the fact that inside this region, the estimated flux is governed by the regularization term, which is described by a single regularization parameter. Therefore, the values inside these regions depend little on the uncertainties of the original IBEX map. We thus need to use a different method to estimate these uncertainties. For this purpose, we assume that the global variance of the GDF flux inside the gaps and ribbon mask is the same as outside of the mask. Therefore, we construct a covariance matrix $\mathbf{V}'_{\mathrm{G}}$ in which the entries corresponding to the data points in the gaps or ribbon mask are replaced with the variance of the observed fluxes over the rest of the sky. Similarly, we define $\mathbf{V}'_{\mathrm{T}}$ where only the data gaps variances are changed. Let $\mathbf{M}'_{\mathrm{T}}$ and $\mathbf{M}'_{\mathrm{G}}$ denote matrices as defined in Equation (10–11), except matrices $\mathbf{V}'_{\mathrm{T}}$ and $\mathbf{V}'_{\mathrm{G}}$ replace matrices $\mathbf{V}_{\mathrm{T}}$ and $\mathbf{V}_{\mathrm{G}}$, respectively. With these definitions, the uncertainties of the spherical harmonic coefficients for the total and GDF map can be calculated as follows:

$$\mathbf{W}'_{a} = \mathbf{M}'_{a}\mathbf{V}'_{a}(\mathbf{M}'_{a})^{\mathrm{T}} \tag{16}$$

This equation does not apply to the ribbon uncertainties because the calculations of the total and GDF coefficients use different uncertainties for pixels inside the ribbon mask. To appropriately account for the uncertainty of the ribbon mask, we need to define a covariance, including both possibilities in the ribbon mask. Let $\mathbf{V}''$ denote a diagonal matrix constructed by expanding matrix $\mathbf{V}'_{\mathrm{T}}$ by $N_{\mathrm{mask}}$ columns and rows, where $N_{\mathrm{mask}}$ is the number of pixels in the ribbon mask. Then, on the diagonal of the expanded part, we put the variance of the observed ENA fluxes from outside the ribbon mask.

Matrices $\mathbf{M}_{\mathrm{T}}$ and $\mathbf{M}_{\mathrm{G}}$ need to be redefined with the new dimensionality of the covariance matrix. In both matrices, the covariance matrices are replaced with one common matrix $\mathbf{V}''$. Additionally, in the definition of matrix $\mathbf{M}_{\mathrm{T}}$ (Equation (10)), we extend matrix $\mathbf{Y}$ by adding $N_{\mathrm{mask}}$ zero vectors as additional columns of this matrix. For matrix $\mathbf{M}_{\mathrm{G}}$, matrix $\mathbf{Y}$ is changed by moving columns corresponding to the pixels to the end of the matrix, while the values at the original columns are replaced with zeroes. Let $\mathbf{M}''_{\mathrm{T}}$ and $\mathbf{M}''_{\mathrm{G}}$ denote these modified matrices. The spherical harmonic representation uncertainty of the ribbon is given by the covariance matrix:

$$\mathbf{W}''_{\mathrm{R}} = (\mathbf{M}''_{\mathrm{C}} - \mathbf{M}''_{\mathrm{G}})\mathbf{V}''(\mathbf{M}''_{\mathrm{C}} - \mathbf{M}''_{\mathrm{G}})^{\mathrm{T}}. \tag{17}$$

We reconstruct the covariance matrix in the standard IBEX pixelization from Equations (16–17) using Equation (15). The square root of the diagonal values is presented in the bottom row of Figure 5, which shows that the newly obtained uncertainties are significantly higher in the ribbon mask and data gap regions. These uncertainties are reported in the derivative product release in connection with this paper.

## 3. Results

We apply the methodology presented in Section 2 to the Compton-Getting and survival probability corrected ram-only IBEX maps from IBEX Data Release #16. This data release includes observations over a full solar cycle from 2009 through 2019. We use the corresponding mask obtained for each energy step, as discussed in Section 2.4. The spherical harmonic representation obtained from our analysis is used to reconstruct IBEX maps with 6°×6° pixelization separately for the GDF and ribbon components.

The GDF-only maps are shown in Figure 6. We adjust the color scale for each energy step to represent the ENA flux range over the considered period. The maps show that the strongest temporal evolution of the GDF is observed in the highest energy step, while the evolution of the GDF in the energy step 0.7 keV is very weak. Furthermore, the enhancement near the heliospheric nose, which is visible in the center of the presented maps, shows the importance of the separation because this enhancement overlaps the ribbon region. The recent enhancement in observed ENA fluxes near this enhancement, which follows the increase in the solar wind dynamic pressure (McComas et al. 2019; Zirnstein et al. 2022), is clearly visible in separated maps from observations in 2017 and later, especially in the higher energy steps.

However, small-scale variations observed in single-year maps need to be carefully considered, as they may result from statistical uncertainties of the original maps (see Section 2.3). The lowest energy step appears as the most spatially dynamic in Figure 6, but only low-degree spherical harmonics are needed to reproduce most of the GDF as discussed in Section 4.2. Consequently, the variations visible in these maps are due to statistical noise and do not show real small-scale structures. Similar variations are also present in the source IBEX maps but are less apparent because the standard IBEX maps are rendered using a perceptually not-uniform "rainbow" color scale, and their color range is broader to include the sum of the GDF and ribbon flux. Moreover, some of these structures appear aligned with the ribbon mask region. The regularization term suppresses high-degree spherical harmonic coefficients. However, the spherical harmonic degree is related to the combined angular scale of structures. Therefore, structures aligned with the mask are less constrained.

Figure 7 presents the ribbon flux maps. Similar to the GDF, the lowest energy step does not show significant time evolution, but time changes are more prominent in higher energy steps. The three lowest energy steps have a clear ribbon structure over the entire solar cycle period, but the energy steps 2.7 and 4.3 keV show that the ribbon observed in recent years is weaker than in the early years of the mission. The employed color scale in the maps makes small changes visible even though most are not statistically significant. The time-combined maps show fewer small-scale structures than those presenting results for individual years. Most visible variations are caused by the statistical noise present in the original IBEX maps. While the high-degree spherical harmonics are needed to reproduce the ribbon profile, they also result in the reproduction of the statistical noise. The IBEX observations do not provide sufficient statistics needed to resolve the small-scale variations predicted by some IBEX ribbon models (Giacalone & Jokipii 2015; Zirnstein et al. 2020). In addition, the ribbon structure in energy step 4.3 keV appears less coherent because the ribbon flux is very low, and for the adopted color scheme for these maps, the statistical noise become visible in this energy step. The GDF reconstructions do not account for the statistical noise inside the ribbon mask. Therefore, the ribbon maps include the total statistical noise present in the original IBEX maps from both the GDF and ribbon components.

Unlike the GDF maps, the ribbon maps do not clearly indicate "recovery" of the ribbon in the last three years. The brightening of the ribbon in the highest energy step does not have a ribbon-like structure and is likely caused by the increase in the GDF over this region. Some enhancement is present in energy steps 1.1, 1.7, and 2.7 keV in years 2017-2019, but it is generally weaker than the one observed in the GDF. On the other hand, the northernmost portion of the ribbon (in the left upper part of each map) weakens universally in energy steps 1.7, 2.7, and 4.3 keV without any signs of recovery, which suggest that this part of the ribbon does not recover. A more detailed analysis of the ribbon evolution in 2009 vs. 2019 can be found in Dayeh et al. (2023b).

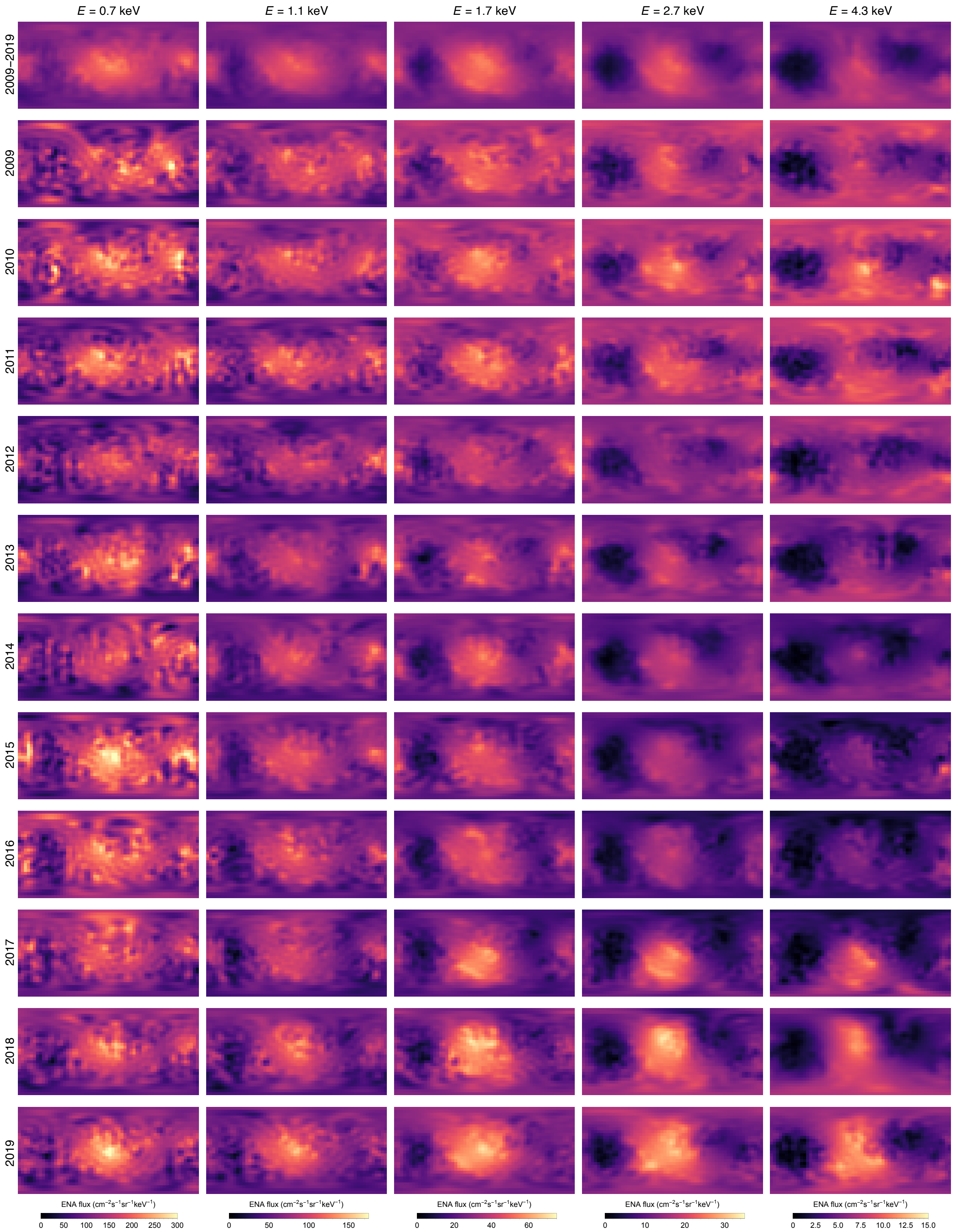

**Figure 6.** GDF maps reconstructed from the spherical harmonic representation for energy steps 0.7, 1.1, 1.7, 2.7, and 4.3 keV (left to right panels). Rows from top to bottom show the results from the time-combined maps and single-year maps from 2009 to 2019.

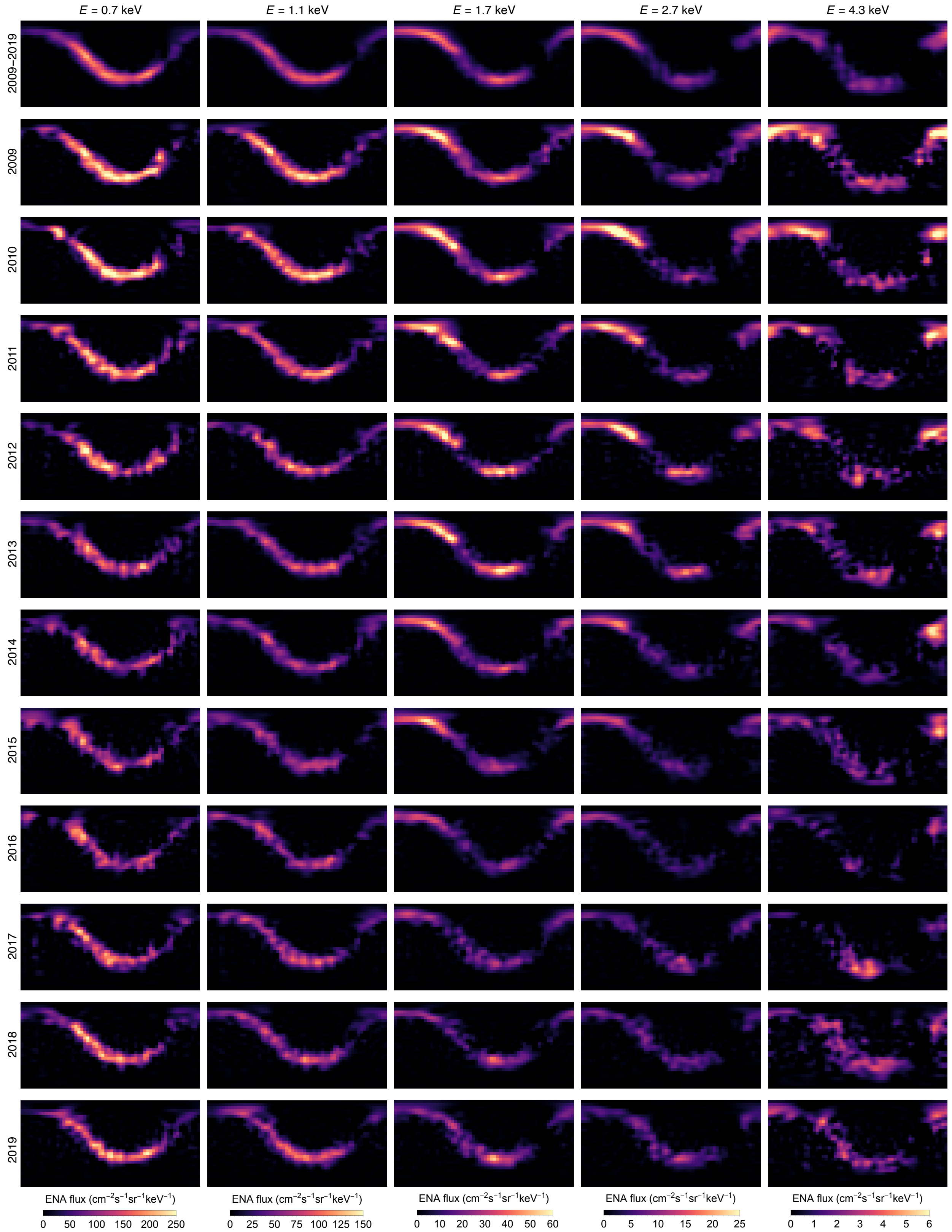

**Figure 7.** Ribbon maps reconstructed from the spherical harmonic representation for energy steps 0.7, 1.1, 1.7, 2.7, and 4.3 keV (left to right panels). Rows from top to bottom show the results from the time-combined maps and single-year maps from 2009 to 2019.

The time-evolution analyses of the GDF in previous studies often avoided the regions overlapping with the ribbon. Utilizing our methodology, we can analyze the time series of each component in any desired direction in the sky. Here, we calculate the time series of the GDF and ribbon flux averaged over the entire sky and six regions centered at the heliospheric nose, tail, north and south ecliptic poles, and starboard and port sides. We average over a portion of the sky within 20° from each direction. Appendix B presents tools allowing for transformation from the spherical harmonic coefficients to averages over any predefined region of the sky. The position of the nose is at ecliptic (255.59°, 5.14°) (Swaczyna et al. 2022b), the tail direction is antipodal to the nose, and the port and starboard are centered at points in the ecliptic plane and 90° away from the nose. We note that we scale the ribbon flux by a factor of 4 in Figure 8 to make visual comparison clearer.

The time series of the GDF and ribbon flux over the entire sky (left column in Figure 8) confirm that the GDF undergoes substantial evolution in the higher energy steps, with evident recovery in the most recent years. In contrast, the evolution of the ribbon is weaker. The GDF increase following the solar wind dynamic pressure increase is the strongest in the nose direction, but it is not visible in the tail direction, in which the GDF continues to decline. Moreover, while the initial maximum of the GDF in the nose is observed in most energy steps in 2010, it is delayed by ~2 years in the tail direction, particularly at 1.1 and 1.7 keV. Thus, the recent solar wind dynamic pressure increase may be reflected in the tail direction in the coming years. The north and south poles have the best statistics of the ENA flux observations because they are observed near continuously throughout the year. The recent increase is clearly visible in the south and north ecliptic poles but appears ~1 year later in the north pole, which is consistent with other studies of the ENA evolution (e.g., Reisenfeld et al. 2016, 2019; McComas et al. 2020) showing that the ENA source in the north is further away than in the south. On the other hand, the starboard and port sides show statistically similar increases in the last two years, although previous studies have hinted at asymmetric evolution of the flanks (Zirnstein et al. 2017).

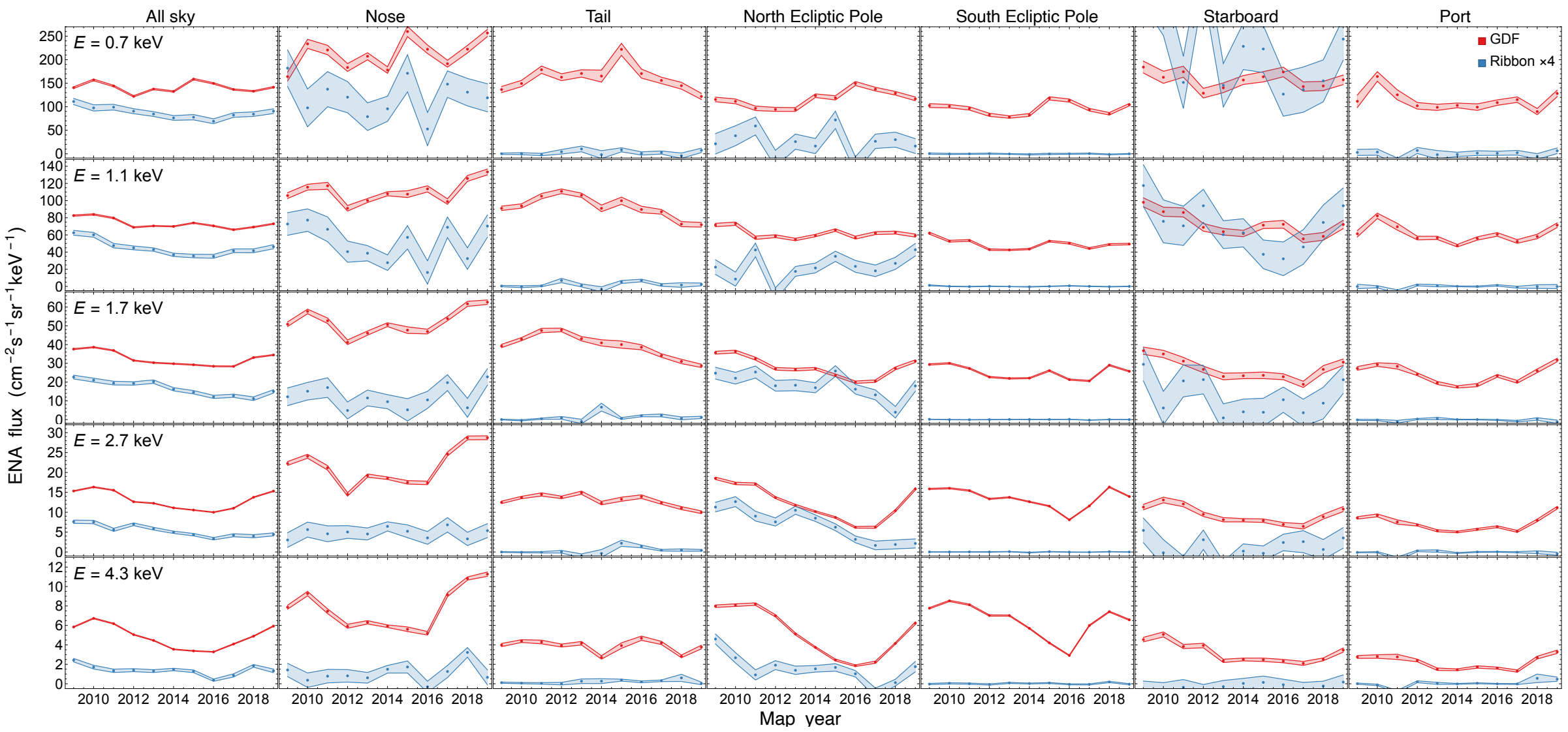

**Figure 8.** Grid of time-series plots of the GDF and ribbon flux observed from the entire sky and six regions around the nose, tail, north and south ecliptic pole, starboard, and port directions (left to right). Rows show results for energy steps 0.7, 1.1, 1.7, 2.7, and 4.3 keV (top to bottom). The ribbon flux is multiplied by 4 to facilitate comparison with the GDF. The bands in this figure show 1σ uncertainties.

The temporal evolution of the ribbon may be tracked from the ribbon-separated maps rotated to the ribbon-centered frame. For comparison between different years, we use the ribbon centers obtained in Section 2.4. The left column in Figure 9 shows polar maps of the ribbon flux averaged over 2009-2011 in the ribbon coordinates. The heliospheric nose lies along the 0° azimuth angle line. For each map, we plot a circle that approximately follows the peak of the ribbon. We integrate the ribbon profiles over the polar angles to analyze the time evolution of the ribbon intensity. The results are shown in the right column of Figure 9. The heliographic latitude of the ribbon's circle for each energy is shown as the top scale. We combine yearly maps similarly as in Zirnstein et al. (2023) into the following year ranges: 2009-2011, 2012-2013, 2014-2015, 2016-2017, and 2018-2019. This combination of fluxes reduces uncertainties and eases the analysis of the solar cycle evolution. Similar combinations have also been used in several studies for similar purposes of examining temporal ENA variations (e.g., Zirnstein et al. 2017; Dayeh et al. 2019, 2022; Schwadron & McComas 2019; McComas et al. 2020).

The obtained profiles for the two lowest energy steps (0.7 and 1.1 keV) show little evolution over time. While the profile for the first three years is slightly higher, the later evolution is mainly within their uncertainties. For the energy step 1.7 keV, the ribbon splits into two main regions: the southern region centered near azimuth +60°, and the northern region centered near azimuth –90°. The southern region indicates some evolution, with a minimum in 2016-2017. The northern region evolves more strongly. The first two periods show similar fluxes, but later the flux starts to decrease; the regions closer to the nose appear to decline earlier than those further away from the nose, which suggests that the ribbon source may be further away, and thus the ENA response to solar cycle changes is delayed. The evolution at 2.7 keV is similar, except the southern region shows a clear maximum in 2012-2013. In the highest energy step, the southern region is weaker than the northern region and remains stable over time. We note that the northern region extends farther away from the nose in this energy step. Additionally, all energies show that the ribbon flux is consistently small over time near azimuth of ~150°. This region is near the port side, i.e., close to the heliospheric tail.

## 1. Discussion

The main motivation for the separation is to allow for separate studies of the GDF and ribbon flux, which evolves differently over the solar cycle, as discussed in the previous section. Previously, studies concerning the GDF often excluded the region containing the ribbon (e.g., Desai et al. 2019; Zirnstein et al. 2021b). With the GDF-only maps, they can be expanded to cover the entire sky. Furthermore, the ribbon-only maps may be used to track the temporal evolution of different ribbon portions. Figure 9 shows that there are regions in which this evolution is statistically significant, while in others, it is minimal. The spherical harmonic decomposition may also be used to study the ribbon profile shape, which is an important characteristic, as different models predict different shapes (Zirnstein et al. 2019a).

The separation of the GDF and ribbon flux in ENA maps enables a further discussion on the angular scale of structures visible in each component. Section 4.1 discusses the power spectra as a function of the spherical harmonic degree for these components in the analyzed energy steps. Later, we continue the discussion of possible small-scale structures in the GDF (Section 4.2). Finally, we discuss the ribbon's geometry reflected in the position of the ribbon's center based on the results obtained from this study (Section 4.3).

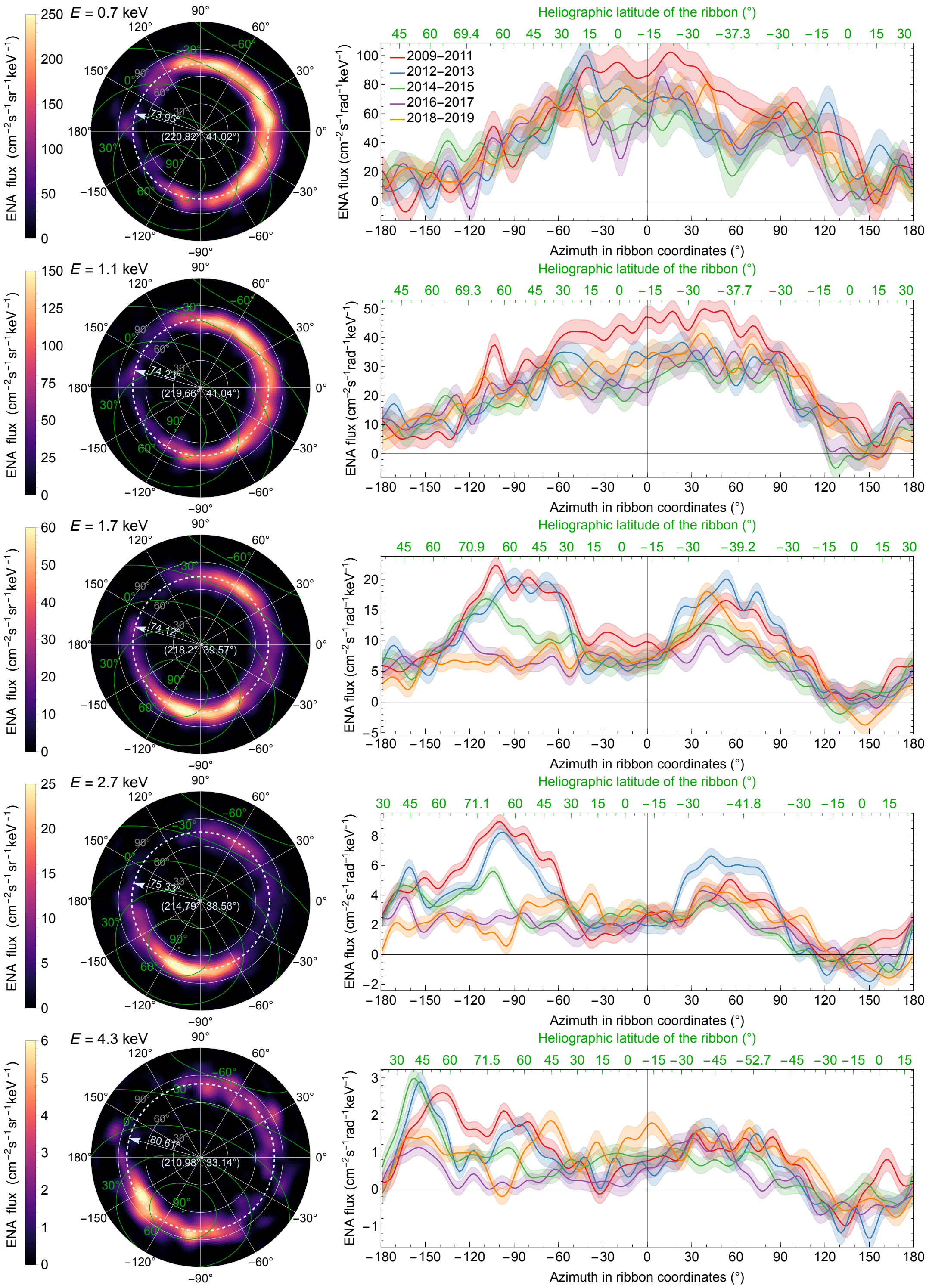

**Figure 9.** Time-evolution of the ribbon flux. *Left column*: polar maps of the ribbon flux averaged over 2009-2011. The dashed circle approximately follows the ribbon's peaks. Lines of heliographic latitudes are shown with green lines. *Right column*: evolution of the polar-angle-integrated ribbon flux as a function of the azimuth. Rows from top to bottom correspond to energy steps 0.7, 1.1, 1.7, 2.7, and 4.3 keV. The bands in this figure show 1σ uncertainties.

## 1.1 Power Spectra of ENA Flux Components

Figure 10 presents the power spectrum of the spherical harmonic representations of each component as a function of the spherical harmonic degree obtained from the time-averaged maps for each energy step. The power spectrum is defined for each degree $\ell$ as

$$P_\ell = \frac{1}{2\ell + 1} \sum_{m=-\ell}^{\ell} c_{\ell,m}^2 \,. \tag{18}$$

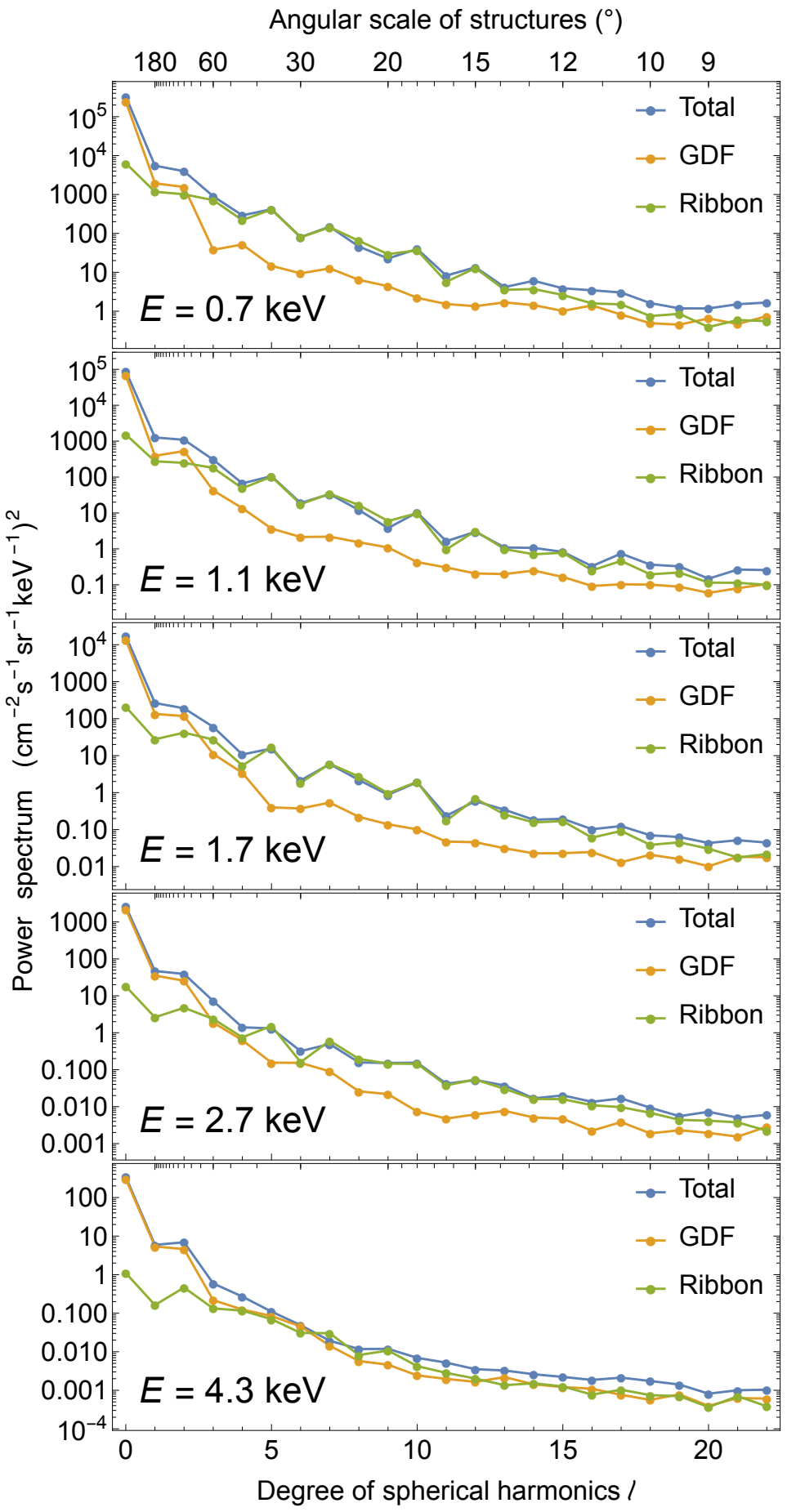

**Figure 10.** Power spectra as a function of the spherical harmonic degree of the GDF and ribbon components and their sum (total) for energy steps 0.7, 1.1, 1.7, 2.7, and 4.3 keV (top to bottom).

The power spectrum of the GDF is larger than the power spectrum of the ribbon component for $\ell \leq 2$ in all observed energy steps. Furthermore, in all energy steps except for the highest, the power spectrum of the ribbon is larger for $3 \leq \ell \leq 15$. In the highest energy step, the power spectra of both components are generally comparable over this range. However, there is a significant range where these two components are comparable, most notably for spherical harmonic degrees between 2 and 4. Therefore, it is impossible to separate the ribbon purely based on the selection of the dominant spherical harmonic components. However, the ratio between the power spectrum of the ribbon and GDF is the largest for the degree of 5, exceeding 20 in the three lowest energy steps. Furthermore, the ratio remains high for these energy steps for several higher degrees. This shows that our methodology successfully separates two signal sources characterized by different angular scales. Visual inspection of Figures 6 and 7 confirms that one of them is the ribbon identified from the first IBEX observations.

The power spectrum of the GDF decreases more significantly for low degrees, confirming that the GDF can be mostly reconstructed from low-degree spherical harmonics. In all energy steps, the GDF power spectrum drops significantly from $\ell = 0$ to 1, from $\ell = 2$ to 3, and for some energies, also from $\ell = 4$ to 5. The constant spherical harmonic describes the mean ENA flux from the entire sky, thus dominating the spectrum. The second drop explains why in the analysis in Paper I, at least spherical harmonics with $\ell \leq 2$ were needed to reconstruct the main structures of the GDF. As discussed in Section 2.4, the structures up to $\ell = 4$ in the GDF should be reconstructed globally, while smaller ones can only be analyzed outside the ribbon mask.

### *1.2 Small-scale Structures in the GDF*

We determine the selection of the maximum degree of spherical harmonics at $\ell_{\mathrm{max}} = 22$ based on the AIC (Section 2.3), from the reconstruction of the total IBEX map, including both the GDF and ribbon components. While the regularization term suppresses the higher degree spherical harmonics, they can reproduce some of the statistical noise visible in the IBEX map, especially those obtained based on a single year of observations. Therefore, we estimate the maximum degree needed to reconstruct the GDF for these maps to identify possible small-scale structures. For this purpose, we calculate the normalized residual sum of squares for each energy step $e$:

$$\chi^2_{\mathrm{GDF},e}(\ell_{\mathrm{GDF},e}) = \sum_t \sum_k \left( \frac{j_{e,t,k} - \sum_{\ell m: \ell \leq \ell_{\mathrm{GDF},e}} y_{k,\ell m} c_{e,t,\ell m}}{\sigma_{e,t,k}} \right)^2 . \tag{19}$$

In the above sum, indices $t$ and $k$ enumerate years and pixels outside the ribbon mask for which observed flux is available in the original IBEX map. We calculate this sum by truncating the reconstruction from the spherical harmonic coefficients to the maximum degree of $\ell_{\mathrm{GDF},e}$.

The sum given in Equation (19) for a normal distribution of the normalized residuals should be approximately equal to the number of data points included in the sum. We chose $\ell_{\mathrm{GDF},e}$ for each energy step so that the sum is the closest to this number. This criterion provides the degrees of 3, 8, 9, 10, and 11 for energy steps 0.7, 1.1, 1.7, 2.7, and 4.3 keV, respectively. The lowest degree is needed for the lowest energy step, mainly because uncertainties are the highest in this energy step. It means that global structures smaller than 180°/3 = 60° are not statistically significant for this energy step. The scale of structures observed in the GDF for higher energy steps decreases to ~20°. This limitation in the angular scale of structures that IBEX observations can resolve is caused by limited statistics of individual IBEX maps. For the time-combined maps, the above criterion would indicate the maximum degrees of 7, 9, 11, 13, and 16 in the respective energy steps. Nevertheless, studies focusing on the GDF-only maps may use the above limited range of spherical harmonics to reconstruct the IBEX maps. Figure 11 shows a version of Figure 6 in which the maximum degree of spherical harmonics included in the reconstruction is limited. This figure shows that a significant portion of the variation visible in Figure 6 disappeared, which confirms that they represented statistical noise. Still, to estimate the ribbon maps, we need higher-degree reconstruction of the GDF to provide an equally good representation of small-scale statistical variations outside of the ribbon mask to reduce the ribbon flux fluctuations outside of the ribbon mask.

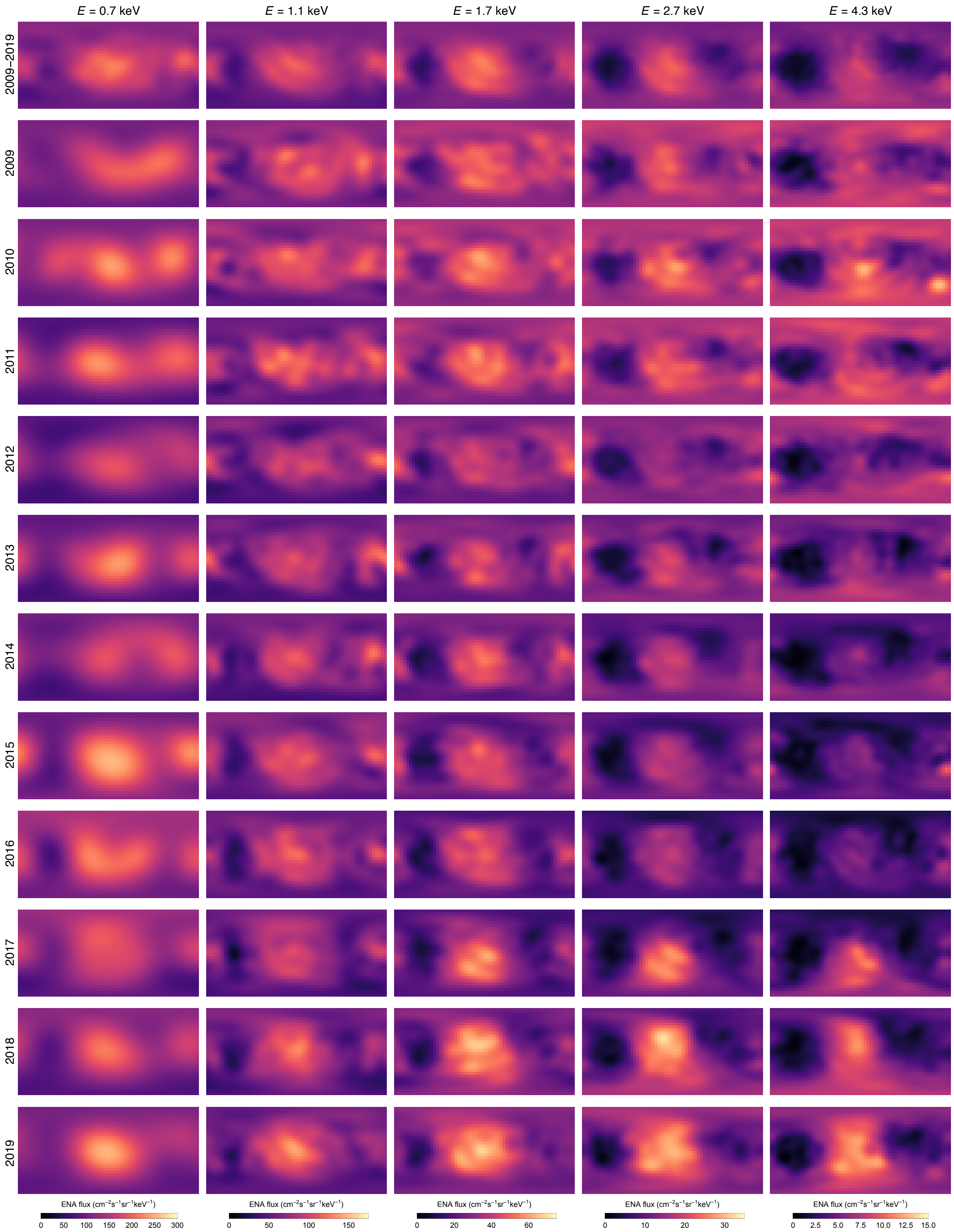

**Figure 11**. As Figure 6 except the GDF flux is reconstructed with the spherical harmonics up to the maximum degree derived in Section 4.2.

It is also interesting to inspect possible outliers in the normalized residuals that may allow for identifying smaller features, including possible point sources. Figure 12 shows histograms of normalized residuals obtained from all single-year maps for the maximum degree found above. The histograms follow the normal distribution, shown with the black line, but some discrepancies require further discussion. Nevertheless, none of the residual signals exceed the 5σ rule used to avoid accidental discovery in comparisons of many data points (e.g., Lyons 2008).

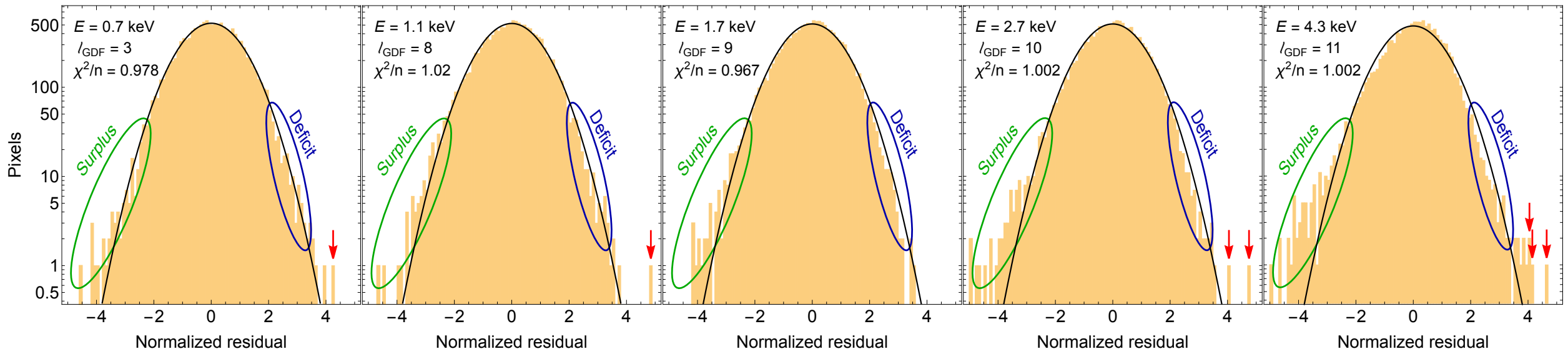

**Figure 12.** Histograms of normalized residual signal between the original IBEX map and truncated GDF estimation (cf. text). Panels from left to right show energy steps 0.7, 1.1, 1.7, 2.7, and 4.3 keV. The histograms are compared with the normal distribution (solid black line). Green and blue ellipses indicate ranges with systematic deviations from the normal distribution, and red arrows point at outliers discussed in the text.

The distribution appears to indicate a surplus of pixels compared to the normal distribution in which the observed flux is smaller than observed, especially for normalized residuals less than about –2 (see green ellipses in Figure 12). At the same time, the histograms are slightly below the normal distribution for normalized residuals between about 2 and 3 (blue ellipses in Figure 12). This effect is likely connected to the estimation of the uncertainty based on the Poisson process. If, for the Poisson process with the true count mean $x$, the observed number of counts $y$ is smaller than $x$, not only is the estimated Poisson process parameter underestimated, but also its standard deviation $\sqrt{y}$. On the other hand, if $y > x$, both the estimated mean and standard deviation are overestimated. Because of this positive correlation, the normalized residuals show an elevated tail for negative normalized residuals.

There are only 8 pixels with positive normalized residuals exceeding 4σ (marked with red arrows in Figure 12). We inspect these pixels in the context of neighboring pixels to verify their importance and to check if they may indicate an interesting signal for further analysis. Six of these pixels are from the two highest energy steps, two from the 2017 map in energy step 2.7 keV, and two in each of the 2017 and 2018 maps in energy step 4.3 keV. All these pixels are centered within 15° from the ecliptic poles and additionally within 63° from the longitude of 180° at which the maps are divided between years. The solid angle covered by pixels within the described limits comprises less than ~2% of the entire sky. As discussed in Paper I, the spherical harmonic representation does not account for time evolution within each year (i.e., as the Earth and thus IBEX orbit the Sun to fill a yearly map); therefore, these regions may not be correctly represented. Furthermore, because the poles are observed throughout the year, the time changes in the ENA flux during the year result in an abrupt spatial change around the longitude at which the map starts and ends each year because the neighboring strips at this longitude are separated by almost a year. Therefore, these six pixels do not represent actual small-scale structures.

The first of the two remaining high residual pixels (4.24σ) is a pixel centered at ecliptic $(\lambda, \beta) = (123°, -39°)$ in the 2012 map for energy step 0.7 keV showing a residual flux of ~92 ENAs $cm^{-2}s^{-1}sr^{-1}keV^{-1}$. Our inspection reveals that there are few neighboring pixels with positive residual fluxes in the same energy step and that this structure is even slightly stronger in the next year, although with a lower statistical significance. Nevertheless, this energy step shows significant time variation, and the pixel strips observed simultaneously show somewhat increased fluxes, suggesting that the background might be underestimated.

The last pixel indicates the most statistically significant positive residual flux of 4.83σ. This pixel, centered at ecliptic $(\lambda, \beta) = (93°, -45°)$, is in the 2017 map at energy step 1.1 keV. The residual flux is ~62 ENAs $cm^{-2}s^{-1}sr^{-1}keV^{-1}$where the reconstructed flux is only ~50 ENAs $cm^{-2}s^{-1}sr^{-1}keV^{-1}$. This residual flux is limited to only this one year, although there is a flux surplus of ~98 ENAs $cm^{-2}s^{-1}sr^{-1}keV^{-1}$ in the same direction at energy step 0.7 keV but at lower statistical significance. The higher energy steps do not show a significant residual flux in this direction. In both energy steps, the positive residuals extend to the two neighboring pixels at the same longitude and to the pixel centered at $(99°, -45°)$. This outlier may therefore be not just a statistical fluctuation but an indication of a compact source of ENAs.

### *1.3 IBEX Ribbon's Center*

The method used in this analysis to find the IBEX ribbon's center is focused on minimizing the width of the region, including the ribbon signal, which is an essential constraint for finding the ribbon mask (Section 2.4). The procedure differs from the two-step fitting technique used in the studies of the ribbon's position (Funsten et al. 2013, 2015; Dayeh et al. 2019). The first fitting is used to find the ribbon's peak position for different azimuthal profiles. In the second fitting, a best-fit circle that follows these peaks is found. Figure 13 compares our method (described in Appendix A) with the previous one (see Appendix in Zirnstein et al. 2023) for the same combined periods discussed in Section 3.

Both techniques reproduce similar patterns of the centers' positions in the observed energy steps. In most cases, the centers from the lowest to highest energy steps are ordered by ecliptic latitude, with the lowest one being closest to the north pole. The ordering along a heliographic meridian relates to the solar wind structure reflected in the IBEX ribbon (Swaczyna et al. 2016b). Moreover, the ordering in 2012-2015 appears to be more closely ordered along the meridian, while it is somewhat tilted in earlier and later years. The ribbon in the energy step 4.3 keV was very weak in 2014-2017 (see also Section 3), and thus our approach failed in finding the ribbon's center. Nevertheless, the two-step fitting can still find a fit for the 2014-2015 map.

There are two systematic differences in the results obtained with these two methods. First, the centers obtained from the two-step fitting are shifted by about +2° in ecliptic longitude. Moreover, the centers in the highest energy step found in this method are shifted south compared to the method from Appendix A. It is important to note that these two methods define the ribbon differently; thus, this difference does not invalidate any of these methods. However, for future comparisons with models, it is critical to use the same technique for both the modeled and observed fluxes.

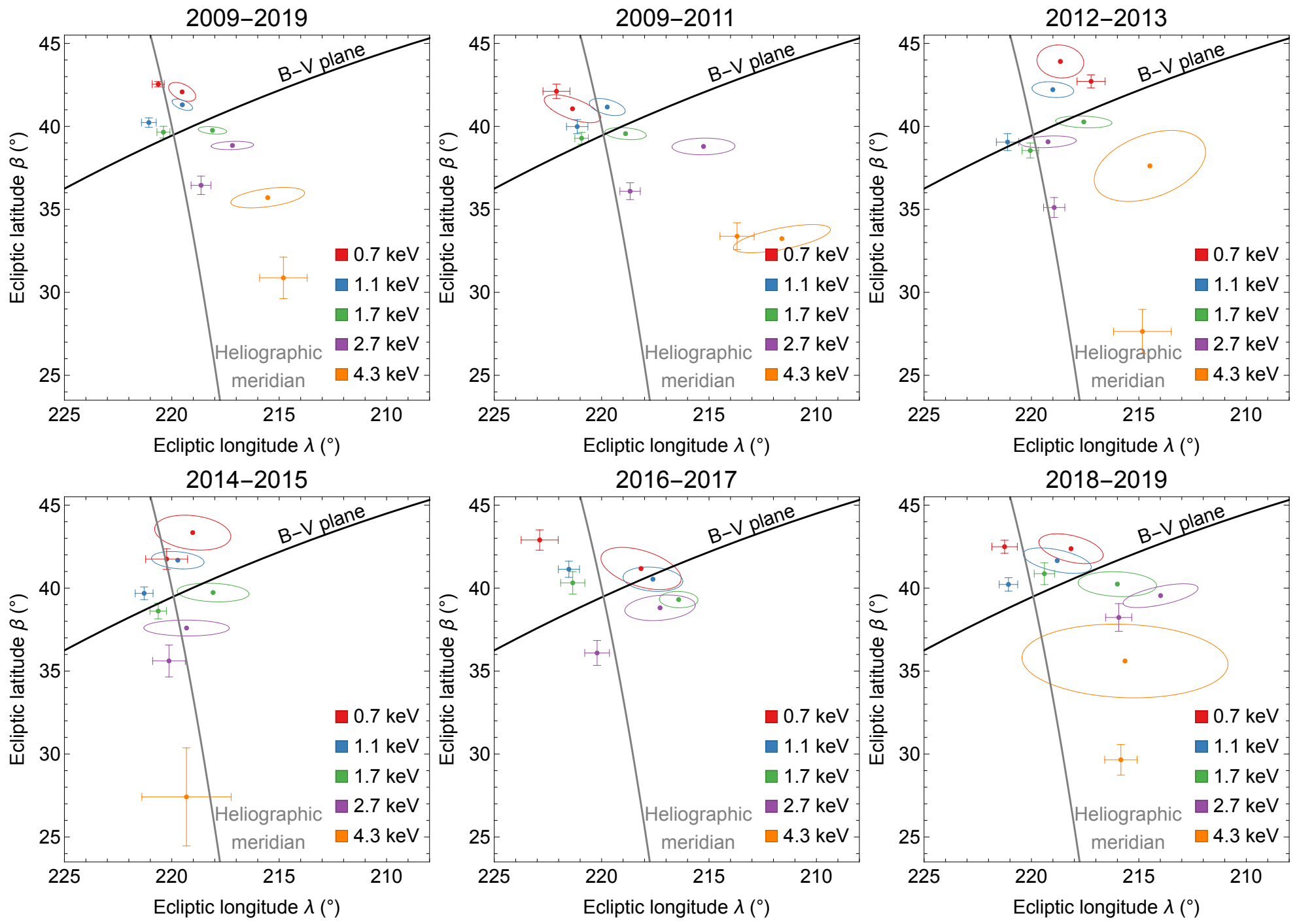

**Figure 13.** Positions of the IBEX ribbon's centers in ecliptic coordinates. The ellipses and points with error bars show the positions obtained using the methodologies presented in Appendix B and by Zirnstein et al. (2023, Appendix), respectively. Panels from the top left to bottom right show results for periods: 2009-2019, 2009-2011, 2012-2013, 2014-2015, 2016-2017, and 2018-2019.

## 2. Summary

Spherical harmonic representations of smoothly changing functions defined on a sphere are a helpful tool for representing it with a finite number of coefficients. The spherical harmonics are ordered by their degree, with higher degrees allowing for the representation of smaller-scale features of the function. The ENA flux maps from IBEX are examples of functions expected to vary smoothly over the sky and may be represented by a finite linear combination of spherical harmonics. Our analysis shows that IBEX maps require spherical harmonics up to the maximum degree $\ell_{\max} = 22$ (Section 2.3), which means that one map is represented by values of $(22 + 1)^2 = 529$ coefficients of spherical harmonics for degrees $\ell < \ell_{\max}$.

The spherical harmonic coefficients are obtained from the least-squares minimization (Section 2.1) supported with a regularization term, suppressing possible artificial extrema in the data gap regions (Section 2.2). This combination allows for independent analyses of single-year IBEX maps. Additionally, it enables the estimation of the GDF component in the data gaps and ribbon region (Section 2.4). Based on this analysis, we provide the spherical harmonic representation of the total (combined) ENA flux, as well as the GDF and ribbon components and their uncertainty matrices (Section 2.5).

Based on the separated signals, we confirm that the GDF and ribbon evolve differently over the solar cycle (Section 3). The GDF shows apparent enhancement following the solar wind dynamic pressure increase in late 2014 (McComas et al. 2019; Zirnstein et al. 2022). The response observed in the ENA flux is delayed, reflecting the distance to the ENA source region. The ribbon flux also evolves over the solar cycle, but it appears to be connected to the evolution of the latitudinal structure of the solar wind over the solar cycle. In contrast to the GDF, the ribbon has not responded strongly to the solar wind dynamic pressure increase.

The degree of a spherical harmonic represents the characteristic spatial scale of structures. Section 4.1 analyzes the power spectra of the ENA flux components as a function of this degree. Low-degree spherical harmonics dominate the GDF, showing that most of the signal smoothly varies over the sky. On the other hand, the ribbon power spectrum is flatter, indicating that the proper representation requires higher degrees of spherical harmonics because the ribbon profile is narrow. While most of the IBEX signal can be represented by spherical harmonic representation, we found that it is currently limited due to temporal changes within each year, which may lead to abrupt changes at the edge of subsequent maps. Additionally, we identified two possible very small ENA sources (see Section 4.2), which require further analysis beyond the scope of this paper. Finally, the new method used to find the ribbon's center in this paper reproduces the most important features of the ribbon geometry, but the found centers do not agree with the previously used two-step method because of the difference in the methods' definitions (Section 4.3).

This paper serves as a detailed description of the spherical harmonic representation of IBEX maps and the separation of the GDF and ribbon components. Concurrently with this paper, we release the obtained spherical harmonic representations on Zenodo under Creative Commons Attribution License (https://doi.org/10.5281/zenodo.7683357). In most situations, this derivative product release should be preferred over the release provided with Paper I. In this release, we include both the spherical harmonic coefficients with their covariance matrices for the total map and the ribbon and GDF components. Additionally, we also provide reconstructed maps in the standard IBEX pixelization. However, unlike the uncertainties in the standard IBEX maps, the corresponding uncertainties are highly correlated, and thus the pixels from these maps should not be statistically combined into a larger region. Instead, the procedure described in Appendix B should be applied in these situations. The methodology developed in this paper may be particularly useful for future analyses of ENA maps from upcoming Interstellar Mapping and Acceleration Probe (IMAP) mission (McComas et al. 2018).

*Acknowledgments*: This material is based upon work supported by the National Aeronautics and Space Administration (NASA) under grant No. 80NSSC21K0582 issued through the Heliophysics Guest Investigators – Open 2020 Program.

## Appendix A. Calculating the IBEX Ribbon's Centers

The IBEX ribbon's centers are essential characteristics of the ribbon geometry in different energy steps. Previous methodologies (Funsten et al. 2013, 2015; Dayeh et al. 2019; Zirnstein et al. 2023) used to find the ribbon's centers adopted a two-step fitting approach in which an IBEX map is rotated to coordinates in which the ribbon's center is close to the north pole. In these ribbon-center coordinates, the ribbon profiles for different azimuths are fit using a Gaussian function with the addition of a linear or quadratic function modeling the GDF(or skew-Gaussian as in Zirnstein et al. 2021a). The peak of the Gaussian described the ribbon position for azimuths where the ribbon is sufficiently strong. In the second step, a circle is fit to the positions of the ribbon peaks.

Separating the ribbon enables a different approach in which we do not need to assume any specific functional form of the ribbon profile. First, we rotate the ribbon spherical harmonic coefficients into possible ribbon centers using the transformation from real to complex spherical harmonics, to which we apply a Wigner D-matrix obtained for the Euler angles describing the rotation. Subsequently, we return to the real spherical harmonic representation, which results in the spherical harmonic coefficients in the rotated frame. This methodology can generally be used on any map represented with spherical harmonics.

For each tested center of the ribbon, we calculate the polar angle profile integrated over all azimuth angles in the rotated frame, from which we find the first and second central moments. For a perfectly Gaussian profile, the first moment indicate the ribbon peak position, while the second central moment represents the ribbon width. In our analysis, we want to find the ribbon center for which the ribbon is as narrow as possible to minimize the ribbon flux outside the mask (Section 2.4). Therefore, we seek the ribbon center for which this second central moment is the smallest. The first moment obtained for this center represents the ribbon radius.

In practice, we calculate the second central moment for possible centers on a rectangular grid in ecliptic coordinates for longitudes from 190° to 250° with a step of 3° and latitudes from 24° to 50° with a step of 2°. We chose slightly larger steps for longitudes to have comparable angular distances in the grid. We find the second central moment for this grid, and we find the minimum using bi-cubic interpolation.

We estimate the center and radius uncertainties using bootstrapping. Based on the covariance matrix of the ribbon spherical harmonic coefficients, we randomly select 100 sets of these coefficients from a multivariate normal distribution described by the best fit ribbon spherical harmonics and their covariance. We repeat the procedure described above, which gives us 100 possible ribbon centers and radii. From this set, we calculate their covariance matrix representing their uncertainties.

## Appendix B. Average Fluxes over Regions from Spherical Harmonic Coefficients

The spherical harmonic representations of ENA flux maps are provided by lists of coefficients and their covariance matrix. However, in most analyses, we are interested in comparisons of the ENA maps with models over some regions of the sky. For example, the average flux over region $\Omega_r$ is given by the following expression:

$$f_r = \frac{\iint_{\Omega_r} \sum_{\ell,m} c_{\ell m} Y_{\ell m}(\theta,\phi) d\Omega}{\iint_{\Omega_r} d\Omega} = \sum_{\ell,m} c_{\ell m} \underbrace{\frac{\iint_{\Omega_r} Y_{\ell m}(\theta,\phi) d\Omega}{\iint_{\Omega_r} d\Omega}}_{z_{r,\ell m}} = \boldsymbol{c} \cdot \boldsymbol{z}_r. \quad \text{(B1)}$$

We change the order of integration and summation because the integrals are finite. Defining vector $\boldsymbol{z}_r = \{z_{r,\ell m}\}_{(\ell m):\ell \le \ell_{\max}}$, this summation can be expressed as a dot product with the vector of the coefficients. Consequently, the integral defining $z_{r,\ell m}$ needs to be only calculated once and later may be applied to multiple maps, e.g., from different years or energy steps. The vectors $\boldsymbol{y}_k$ (Section 2.1) describing the average values over the IBEX pixels are just a special case of this integral, where the regions correspond to IBEX pixels (see also Paper I). Therefore, the discussion provided below applies also to the reconstructed pixelized maps. These integrals can be calculated either numerically or analytically.

Integrals over rectangular ranges of longitudes and latitudes are straightforward. However, this can be combined with a rotation to any other coordinate system. For example, we calculated the average over circular regions centered at different points in Figure 8. For this purpose, we combine the integration of spherical harmonics over the complete azimuth angle and a polar angle from 0 to $r$, where $r$ is the radius of this region with the rotation to the coordinates in which the center of the region coincides with the pole (see Appendix A). A similar procedure may be used to integrate over a ring region or a ring sector.

The vectors $\boldsymbol{z}_r$ for several regions can be combined into a single matrix $\mathbf{Z} = \{\boldsymbol{z}_r\}_{r=1,\dots,N_\mathrm{r}}$, where $N_\mathrm{r}$ is the number of regions. With this definition, the fluxes in all these regions are:

$$\boldsymbol{f} = \mathbf{Z}\boldsymbol{c}. \tag{B2}$$

The covariance matrix of uncertainties in this region is calculated from simple propagation of errors:

$$\boldsymbol{\Sigma}_f = \mathbf{Z}\mathbf{V}\mathbf{Z}^\mathrm{T}. \tag{B3}$$

The same expression is used to calculate the covariance matrix for reconstructing the pixelized maps. It is essential to account for possible correlations when averaging over multiple pixels. Therefore, the procedure in this appendix should be followed for further analyses of various regions that combine pixels because the correlations are significant, especially between neighboring pixels.